\documentclass[journal]{IEEEtranDOI} % Use this version to show DOIs in the references.
\usepackage[english]{babel}
\usepackage[noadjust]{cite}
\usepackage{algorithm}
\usepackage{algpseudocode}
\usepackage{booktabs}
\usepackage[table,x11names]{xcolor}
\usepackage{placeins}
\usepackage{siunitx}
\DeclareSIUnit{\dBm}{\deci\belmilliwatt}
\DeclareSIUnit{\dBi}{\deci\beli}
\usepackage[table]{xcolor} % Make sure to include the 'table' option
\definecolor{lightergray}{rgb}{0.9, 0.9, 0.9}
\usepackage{titlesec}

\ifCLASSINFOpdf
\else
   \usepackage[dvips]{graphicx}
\fi
\usepackage{url}
\usepackage[scaled]{helvet}

\usepackage{multirow}
\usepackage{tabularx}
\usepackage{graphicx}
\usepackage{optidef}
\usepackage{amsmath}
\usepackage{bm}
\usepackage{amssymb}
\usepackage{xcolor}
\usepackage{caption}
\usepackage{subcaption}
\usepackage{booktabs}
\usepackage{lipsum}% 
\usepackage{siunitx}  
\usepackage{float}
  \usepackage{placeins}
  
\usepackage[font=small,labelfont=sc]{caption}
\usepackage[colorinlistoftodos, textsize=footnotesize]{todonotes}

\usepackage[acronym]{glossaries}
\usepackage[hidelinks]{hyperref}
\usepackage{cleveref}
\Crefname{figure}{Fig.}{Figs.} % Fix figure refs for
\usepackage[binary-units]{siunitx}

\definecolor{c1}{rgb}{0.2, 0.4, 0.8}   % Dark blue (worst)
\definecolor{c2}{rgb}{0.45, 0.65, 0.9} % Medium blue (middle)
\definecolor{c3}{rgb}{0.75, 0.88, 1.0} % Light blue (best)

\DeclareSIUnit{\belmilliwatt}{Bm}
\DeclareSIUnit{\dBm}{\deci\belmilliwatt}
\DeclareSIUnit{\belisotropic}{Bi}
\DeclareSIUnit{\dBi}{\deci\belisotropic}
\DeclareSIUnit{\foot}{ft}

\makeglossaries

\newacronym{itm}{ITM}{Irregular Terrain Model}
\newacronym{ehata}{eHata}{extended Hata}
\newacronym{us}{US}{United States}
\newacronym{id}{ID}{identifier}
\newacronym{rsrp}{RSRP}{Reference Signal Received Power}
\newacronym{gps}{GPS}{Global Positioning System}
\newacronym{earfcn}{EARFCN}{E-UTRA Absolute Radio Frequency Channel Number}
\newacronym{lsf}{LSF}{large-scale fading}
\newacronym{aoa}{AoA}{angle of arrival}
\newacronym{aod}{AoD}{angle of departure}
\newacronym{bs}{BS}{base station}
\newacronym{gis}{GIS}{geographic information system}
\newacronym{ml}{ML}{machine learning}
\newacronym{mae}{MAE}{mean absolute error}
\newacronym{rmse}{RMSE}{root mean square error}
\newacronym{los}{LoS}{line-of-sight}
\newacronym{nlos}{NLoS}{non-LoS}
\newacronym{dsm}{DSM}{Digital Surface Model}
\newacronym{dhm}{NDHM}{Normalized Digital Height Model}
\newacronym{qos}{QoS}{quality of service}
\newacronym{smote}{SMOTE}{Synthetic Minority Over-sampling Technique}
\newacronym{ntia}{NTIA}{National Telecommunications and Information Administration}
\newacronym{vbw}{VBW}{vertical beamwidth}
\newacronym{3d}{3D}{3-dimensional}
\newacronym{sui}{SUI}{Stanford University Interim}
\newacronym{spm}{SPM}{Standard Propagation Model}
\newacronym{ue}{UE}{user equipment}
\newacronym{3gpp}{3GPP}{3rd Generation Partnership Project}
\newacronym{ai}{AI}{Artificial intelligence}
\newacronym{snr}{SNR}{signal-to-noise ratio}

\usepackage{afterpage}

\usepackage{titlesec}
\titlespacing\section{0pt}{*1}{*1}
\titlespacing\subsection{0pt}{*0.8}{*0.8}
\titlespacing\subsubsection{0pt}{*0.6}{*0.6}

\usepackage{eso-pic}
\usepackage{xcolor}

\newcommand{\IEEEarxivnotice}{%
\AddToShipoutPictureFG*{%
  \AtPageUpperLeft{%
    \raisebox{-0.18in}{%
      \makebox[\paperwidth][c]{%
        \fbox{%
          \begin{minipage}{0.88\textwidth}
          \centering
          \scriptsize
          This work has been submitted to the IEEE for possible publication.
          Copyright may be transferred without notice, after which this version may no longer be accessible.
          \end{minipage}
        }%
      }%
    }%
  }%
}%
}

\begin{document}
\IEEEarxivnotice

\bstctlcite{BSTcontrol}

\title{Simulation-Driven Ensemble Machine Learning for Robust and Generalizable Path Loss Prediction}

\author{
Ahmed~P.~Mohamed,~\IEEEmembership{Graduate Student Member,~IEEE,}~% This stops the extra space.
Byunghyun~Lee,~\IEEEmembership{Graduate Student Member,~IEEE,}~% \\
Yaguang~Zhang,~\IEEEmembership{Member,~IEEE,}~%
Christopher~R.~Anderson,~\IEEEmembership{Senior~Member,~IEEE,}
~David~J.~Love,~\IEEEmembership{Fellow,~IEEE,}~% \\
and James~V.~Krogmeier,~\IEEEmembership{Senior~Member,~IEEE}

\thanks{A. P. Mohamed, B. Lee, D. J. Love, and J. V. Krogmeier are with the Elmore Family School of Electrical and Computer Engineering, Purdue University, West Lafayette, IN 47907 USA (e-mail: \{mohame23, lee4093, djlove, jvk\}@purdue.edu).}%
\thanks{Y. Zhang is with the Department of Agricultural and Biological Engineering, Purdue University, West Lafayette, IN 47907 USA (e-mail: ygzhang@purdue.edu).}
\thanks{C. R. Anderson is with the NTIA Institute for Telecommunication Sciences, Boulder, CO 80305 USA (e-mail: canderson@ntia.gov).}
\thanks{This work is supported by the National Science Foundation under grants EEC-1941529, CNS-2212565, and CNS-2225578. Preliminary results related to this paper appeared in IEEE ICC 2024~\cite{mohamed2024simulation}.}
% https://journals.ieeeauthorcenter.ieee.org/wp-content/uploads/sites/7/IEEE-Editorial-Style-Manual-for-Authors.pdf
}

\maketitle
% \IEEEarxivnotice
% \vspace*{0.1in}

\begin{abstract}
Machine learning has emerged as a promising approach to path loss prediction, yet its effectiveness often degrades when measurement data are scarce. To address this limitation, we propose an ensemble-based machine learning framework that integrates real-world measurements with synthetic data generated using a lidar-based simulator. The simulator provides broad spatial coverage at low cost by producing static path loss values that capture terrain variations and physical obstacles present in the propagation environment. A dynamically weighted ensemble then combines the simulation results with measured data to balance the contribution of both data sources and improves generalization across diverse environments. To further mitigate the effects of limited measurements over selected environments, we incorporate the \gls{smote}, a data augmentation technique that synthesizes additional samples through interpolation between existing measurements while preserving their statistical properties. By leveraging simulation data, SMOTE, and carefully engineered propagation features as inputs, the proposed framework captures both geographical and physical variability, enabling adaptability across urban, suburban, residential, industrial, and  rural environments. Experimental results demonstrate that the proposed method achieves up to a 50\% reduction in \gls{mae} for path loss predictions in dB, compared with models trained solely on real data, and up to a 25\% improvement relative to models trained exclusively on synthetic data, particularly for cross-environment generalization. These findings highlight the effectiveness of combining simulation-based synthetic data with SMOTE to overcome data scarcity and enhance the model's generalization ability. Overall, the proposed framework provides a robust and practical solution for trustworthy path loss prediction across diverse environments with limited measurement data, supporting cost-effective planning and optimization of next-generation wireless networks.

\begin{IEEEkeywords}
Data augmentation, geographical information features, machine learning, path loss, site-specific wireless propagation.
\end{IEEEkeywords}

\end{abstract}

\IEEEpeerreviewmaketitle

\section{Introduction}

\glsresetall

Ubiquitous wireless broadband---essential to modern economic activity---continues to drive demand for both additional spectrum and greater bandwidth within allocated bands, resulting in ever-increasing congestion. Efficient spectrum management for future wireless systems requires accurate and reliable models to optimize use of crowded spectrum.

Specifically, propagation models for future wireless systems must be developed to support coverage prediction, interference analysis, and mobility management. Yet, developing propagation models is challenging because of the complex interactions between the signal and its environment, including reflection, scattering, diffraction, and transmission through diverse, irregular structures~\cite{phillipsSurveyWirelessPath2013}. Although existing site-general models, e.g., \gls{ehata} the \gls{3gpp} ~\cite{3gpp.38.901}, have proven useful, they were developed with limited measurement data and without access to modern data science tools or detailed site-specific information about the physical environment. 

\gls{ai}/\gls{ml}
 based propagation models have demonstrated improved performance relative to deterministic and empirical models; however, they typically suffer from limited training data, a lack of data diversity, and a loss of the tie to the underlying physics of propagation.  In this work, we explore hybrid models that combine \gls{ai}/\gls{ml} approaches with physics-based propagation models so that the resulting hybrid models provide a more accurate, repeatable, and generalizable model than either methodology would alone.

\subsection{Related Work}
Propagation models generally fall into one of three broad categories: deterministic, semi-empirical, or fully empirical~\cite{ayadi2015two}. A fourth category comprises hybrid approaches that, for example, combine deterministic ray tracing with statistical clutter factors~\cite{kozma2023proposed,zhang2019propagation}.  Deterministic models, such as ray-tracing or parabolic wave equation methods, utilize numerical methods to solve Maxwell's equations directly.  Although these models provide a high degree of prediction accuracy, they require extremely high environmental fidelity, computational complexity, and computational resources.  Semi-empirical models, such as the \gls{itm} ~\cite{huffordDEPARTMENTCOMMERCE} and \gls{ehata} models combine physics-based approximations to Maxwell's equations with empirically derived corrections for a variety of path-based topographical and environmental features.  Fully empirical models, such as COST 231~\cite{COST231}, the 3GPP spatial channel model, and QuadRiGa~\cite{jaeckelQuaDRiGa3DMultiCell2014} incorporate one or more random variables to represent the inherent variability in wireless links, such as shadowing, fading, and scattering.  Semi- and fully empirical models provide broad applicability and computational efficiency; however they suffer from an oversimplification of path- and site-specific features, reducing their accuracy.

Previous work~\cite{thrane2020deep,zhang2022artificial,masood2022interpretable} proposed learning-based models to model propagation effects caused by path- and site-specific features.
Although these models offer the potential to adapt to complex environments and improve prediction accuracy, they require extensive, high-quality datasets and complex preprocessing or location-specific training, limiting their generalizability and scalability in data-scarce or diverse environments.
The collection of such datasets is a demanding task, requiring significant time, effort, and expensive measurement tools. Without sufficient data, learning-based models' accuracy degrades when those models are applied to environments that differ even slightly from those represented in the training set~\cite{vanleer2021improving}. Rural areas, where diverse propagation paths exist over wide, varied landscapes, pose a critical challenge for developing accurate and wireless propagation models~\cite{zhangChallengesOpportunitiesFuture2021a}.

Data augmentation has long been recognized as an effective means to improve the robustness and precision of \gls{ml} prediction models that have limited datasets~\cite{kortylewski2019analyzing}.  
Despite its success in other fields, the application of data augmentation to develop path loss prediction models remains relatively underexplored.
For example, previous work has utilized simulation data to address coverage gaps in inaccessible geographic regions for site-specific channel modeling~\cite{zhangSimulationAidedMeasurementBasedChannel2020}. However, this approach was limited to augmenting a partial dataset for a single geographic area, rather than extending the dataset to cover multiple environments. Similarly, another study incorporated synthetic samples to extend downlink-only datasets and enable prediction for uplink and backhaul scenarios ~\cite{dempsey2025reciprocity}. Although this reduced the uplink prediction error significantly, the method was still confined to a single environment and did not address cross-environment generalization.

A common practice for training \gls{ml} models and one that is prevalent in the literature is to merge synthetic and measurement data into a single training set input to a \gls{ml} model \cite{seib2020mixing}.  Measurement data contains real-world imperfections, such as background noise, hardware nonlinearities, dynamic range limits, or geographic biases.  Synthetic data is often generated from idealized or simplified descriptions of the physical environment and can miss nuances in the physical environment that result in large errors in path loss prediction.  Training a single model on such mixed data risks overfitting the dominant dataset features, learning spurious correlations, or a failing to generalize to new environments~\cite{zhang2015multi,vanleer2021improving}. 
\subsection{Modeling Scheme Overview}
In this paper, we present a \gls{ml} path loss prediction framework that achieves robust performance across diverse geographic cell sites without requiring large measurement datasets. To enhance model generalizability, we introduce a novel data augmentation technique that blends limited real-world measurements with synthetic data generated by a physics-informed propagation model~\cite{zhangLargeScaleCellularCoverage2023}. High-resolution lidar datasets are incorporated to ensure accurate and detailed environmental representations within the propagation model.

To effectively address the varying characteristics of synthetic and real-world data, this work adopts an ensemble learning approach that combines multiple \gls{ml} models, each of which is trained separately on an individual data source, either synthetic or measured. This approach is especially effective in situations where no single dataset fully captures the complexity of the modeled environment. Models trained in isolation can suffer from bias, overfitting, or limited generalization, particularly when data is noisy, imbalanced, or biased. By combining predictions from multiple models, ensemble methods help smooth out these limitations, often leading to more stable and reliable performance~\cite{dietterich2000ensemble, rokach2010ensemble}. 
\begin{figure}[t]
\centering
    \includegraphics[scale=0.18]
    {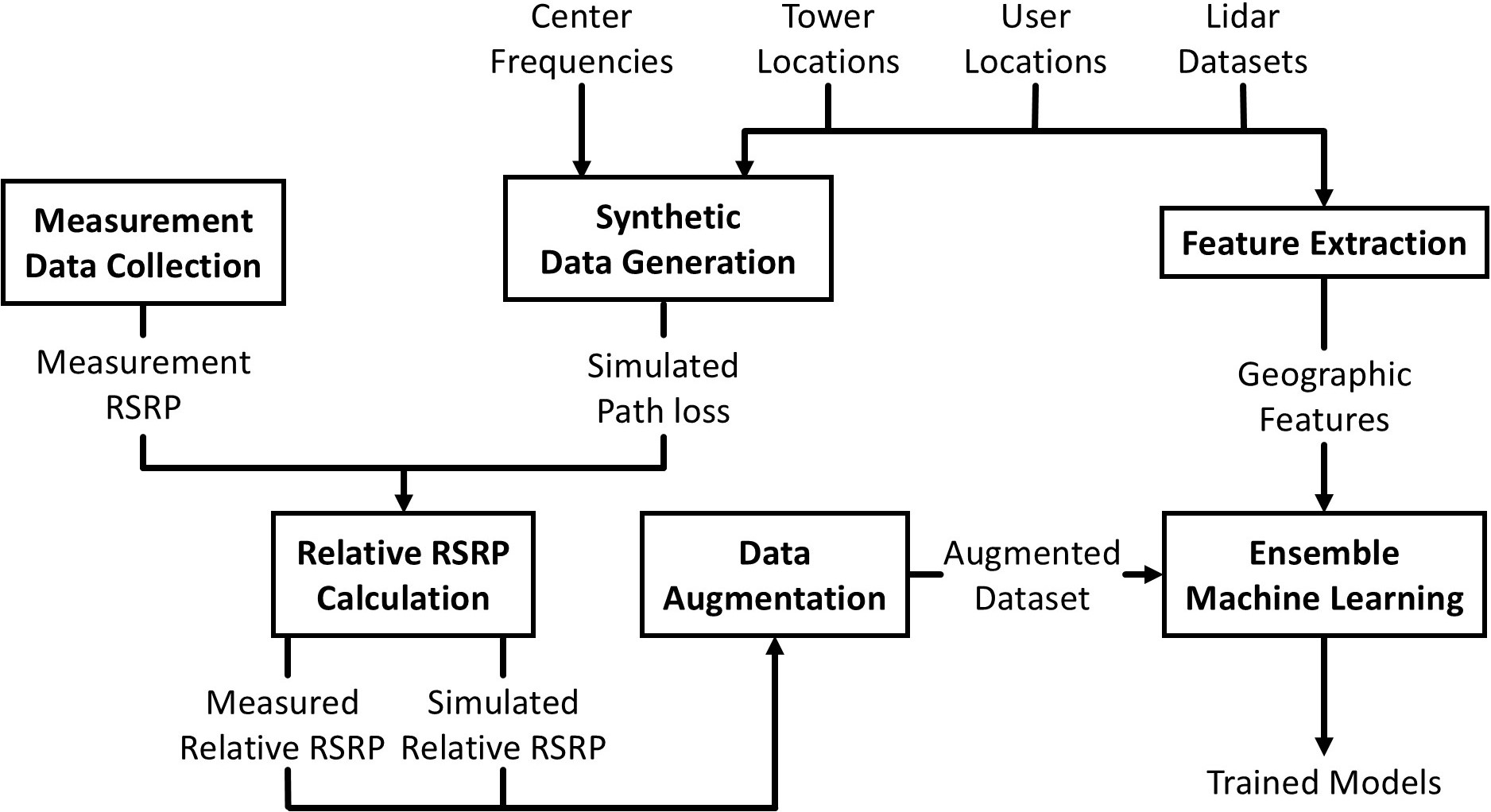}
        \caption{Flowchart of the proposed simulation-driven data augmentation process.}
        \label{fig:flowchart}
\end{figure}
\Cref{fig:flowchart} illustrates the overall process of the proposed data augmentation framework. The three key inputs to the simulation-enhanced data augmentation are:
\begin{itemize}
    \item \textbf{Real-World Measurement Data Collection}: 
    Measurement data collected from a variety of propagation environments. In this work, we utilize RSRP measurements from mobile phones to produce path loss measurements to take advantage of their ability to quickly and inexpensively produce large quantities of propagation measurements.
    \item \textbf{Synthetic Data Generation:} 
    Large-scale simulations configured to address gaps in the measurement datasets, particularly for locations, distances, physical environments, and frequencies that are sparsely represented, thus enriching overall data diversity. 
    \item \textbf{Feature Extraction:}
    Path-specific features extracted from \gls{gis} databases designed to capture the geometric relationships and physical context of each measured propagation link. The use  of these features as a training context for the \gls{ml} models is key to the generalizability of the overall modeling framework.
\end{itemize}

\begin{figure*}[ht]
     \centering
     \begin{subfigure}[b]{0.32\textwidth}
   {\includegraphics[width=.980\linewidth]{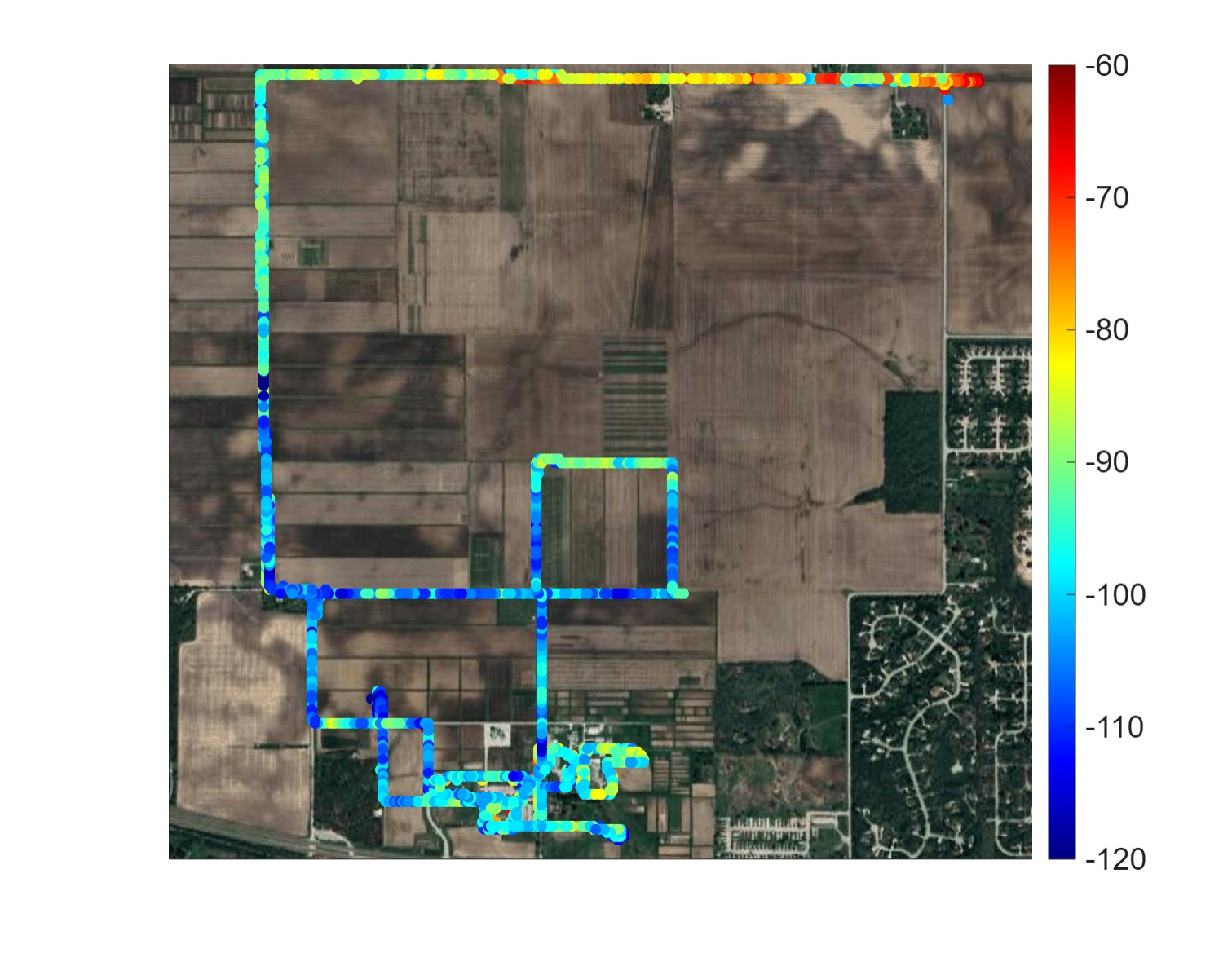}}
   \vspace{-10pt}

            \caption{ Rural (ACRE)}
            \label{fig:ACRE}
     \end{subfigure}
     \begin{subfigure}[b]{0.32\textwidth}
        {\includegraphics[width=.980\linewidth]{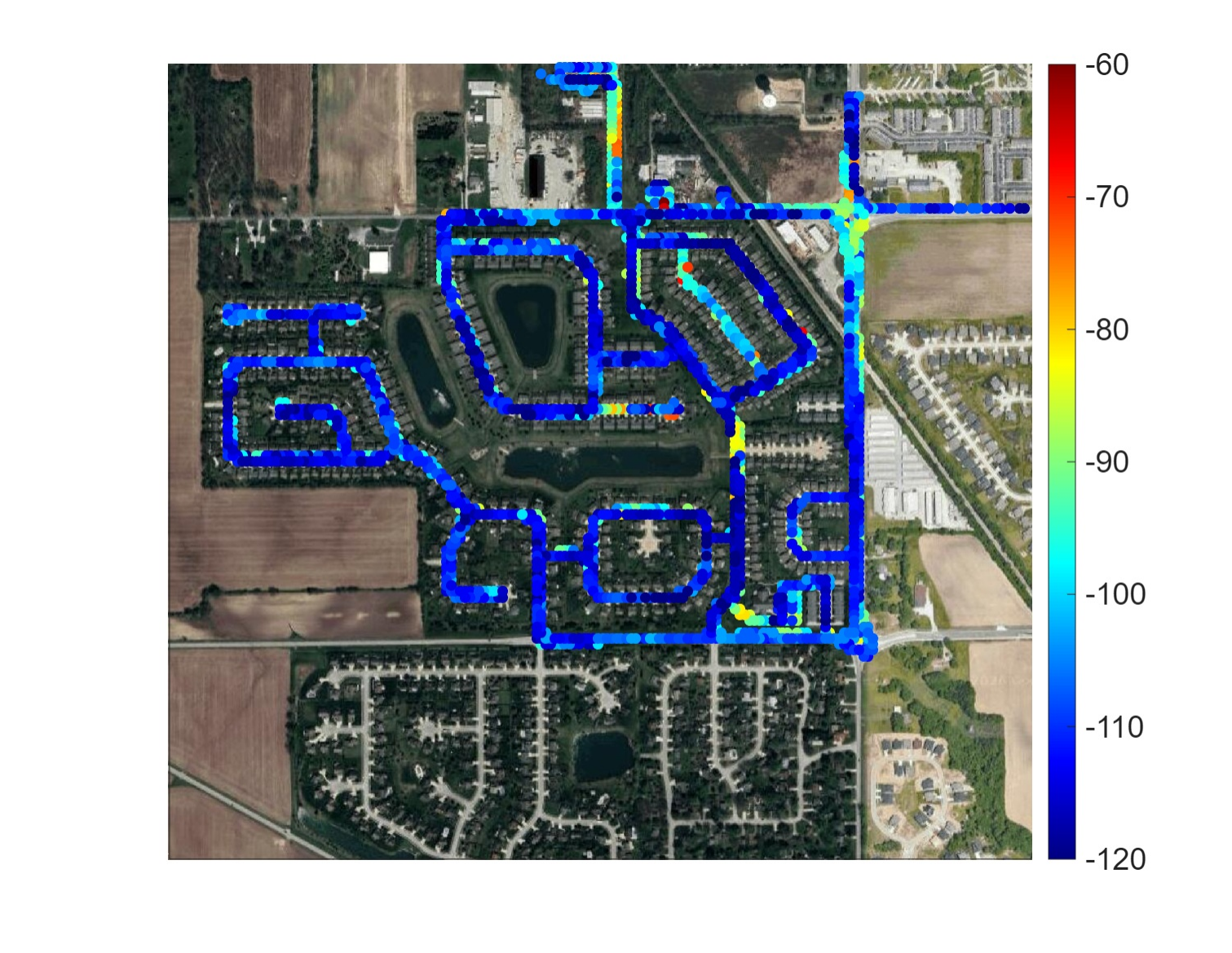}}
           \vspace{-10pt}

             \caption{ Residential (Lindberg)}
             \label{fig:Lindberg}
     \end{subfigure}  
     \begin{subfigure}[b]{0.33\textwidth}
        {\includegraphics[width=.980\linewidth]{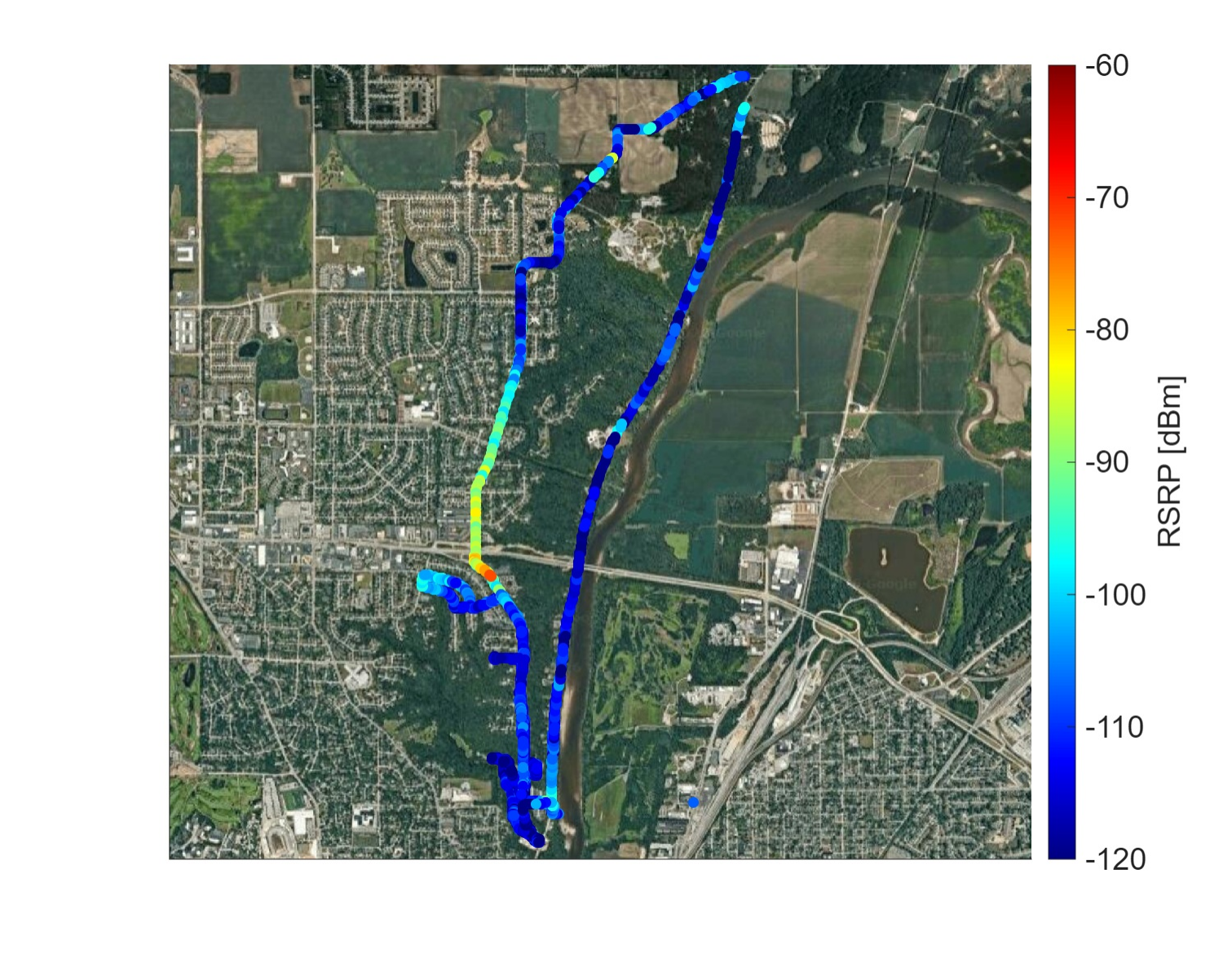}}
           \vspace{-10pt}

             \caption{ Hilly 1 (Happy Hollow)}
             \label{fig:HappyHollows}
     \end{subfigure}   
     \begin{subfigure}[b]{0.33\textwidth}
      \hspace*{-0.8em}
        {\includegraphics[width=.980\linewidth]{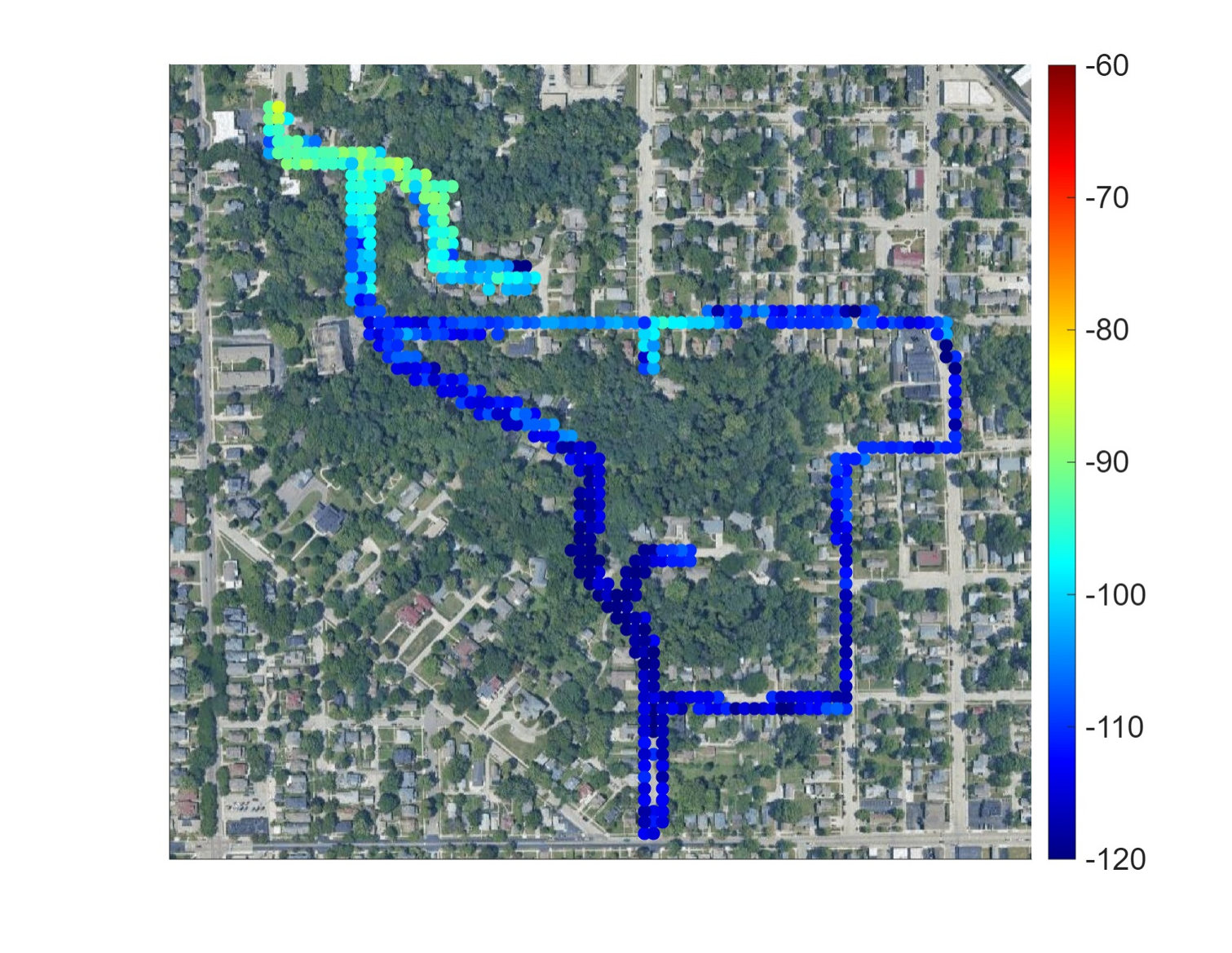}}
           \vspace{-10pt}

             \caption{ Hilly 2 (Valley)}
             \label{fig:valley}
     \end{subfigure}   
\begin{subfigure}{0.30\textwidth}
    \centering
    \raisebox{0.4cm}{%
        \includegraphics[height=3.9cm]{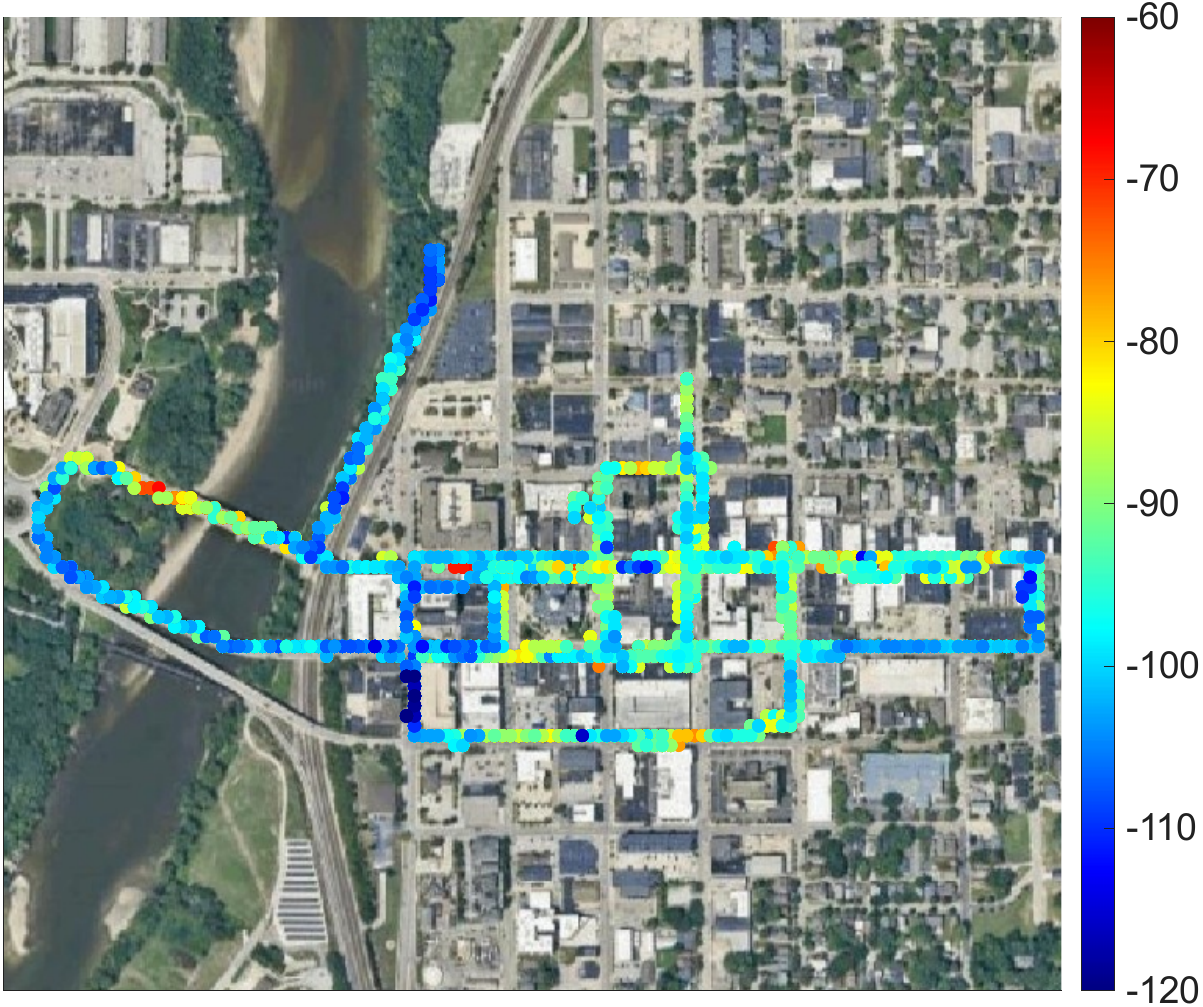}
    }
       \vspace{-10pt}

    \caption{Urban (Downtown Lafayette)}
    \label{fig:downtown}
\end{subfigure}
     \begin{subfigure}[b]{0.325\textwidth}
           % \hspace*{0.2em}
             {\includegraphics[width=.980\linewidth]{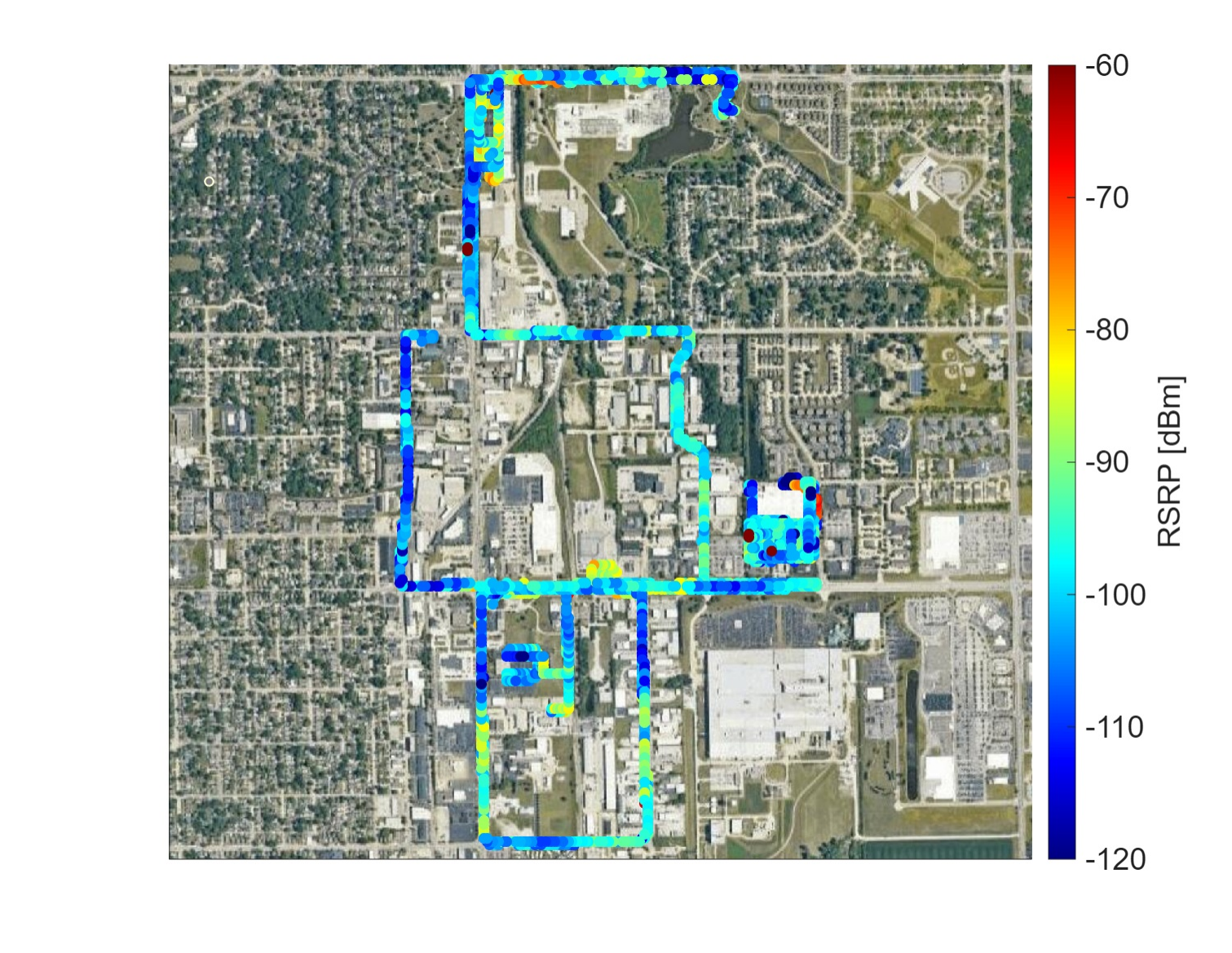}}
                \vspace{-10pt}
             \caption{ Industrial (Southeast Lafayette)}
             \label{fig:target}
     \end{subfigure}   
     \caption{\gls{rsrp} in \si{\dBm} derived from the data collected during the measurement campaign. Satellite imagery: Google Maps.}
     \label{fig:RSRPmap1}
\end{figure*}

Our contributions can be summarized as follows:

\begin{itemize}

\item \textbf{Generalization Across Different Environments:}
The proposed framework demonstrates robust generalizability of the model to new environments that were not included in the training set as validated through an extensive measurement campaign.

\item \textbf{Optimized Training via Data Augmentation and Ensemble Learning:}
Our approach improves overall prediction accuracy by using the \gls{smote} ~\cite{chawla2002smote} to dynamically combine \gls{ml} models trained on synthetic, real, and hybrid datasets into an ensemble model framework.

\item \textbf{Comprehensive Validation and Comparisons:}  
    The developed model framework is validated against both advanced empirical propagation models and measurement data, demonstrating significant improvements in prediction accuracy.
    % These results underscore the practicality and scalability of the proposed approach for real-world applications.
\item \textbf{Common Metric for Unified Real and Synthetic Data:}
We introduce a new metric, Relative \gls{rsrp} ($\Delta\text{\gls{rsrp}}$) to effectively merge measured and synthetic datasets to ensure consistency in training and testing \gls{ml}-based models.
    
\end{itemize}

\section{Measurement and Synthetic Dataset Creation}

\begin{figure*}[ht]
\centering
    \includegraphics[width=\textwidth,keepaspectratio]{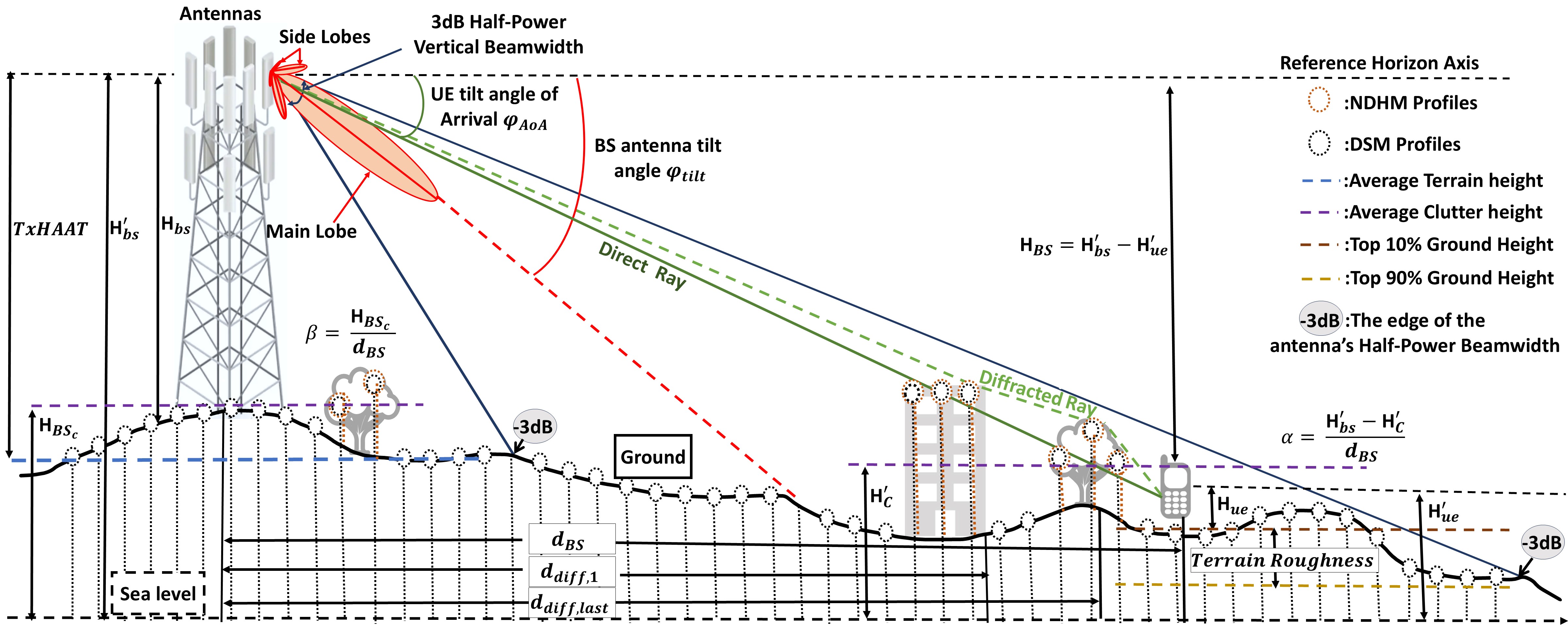}
        \caption{Illustration of the propagation path between a BS and user equipment (UE), highlighting the engineered features used as inputs to the ML algorithm. }
        \label{fig:relativersrpbs}
\end{figure*}

In this section, we describe our procedure for collecting propagation measurement data, generating synthetic data using simulations, and our data postprocessing procedure for ensuring accurate measurement data and merging synthetic data with measurements.

\subsection{Creating Measurement Datasets}
Recent studies have demonstrated that mobile phones can be effectively utilized to gather reliable \gls{rsrp} measurements, thereby facilitating the creation of robust propagation measurement datasets~\cite{hollingsworth2024repurposing, ghosh2025}. We carried out a comprehensive data collection in Tippecanoe County, located in the state of Indiana in the \gls{us}.
We collected data from six different environments: rural, two hilly areas, urban, residential and industrial areas, as illustrated in Fig. \ref{fig:RSRPmap1}.
The 4G LTE measurements were performed using multiple commodity Android devices, including Samsung Galaxy S8, S20, S21, A54, and Oppo phones, with the assistance of the G-NetTrack Pro~\cite{gnettrackApp}, a cellular network monitoring application.\footnote{Certain commercial products or company names are identified here to describe our study adequately. Such identification is not intended to imply recommendation or endorsement by the National Telecommunications and Information Administration, nor is it intended to imply that the products or names identified are necessarily the best available for the purpose.} The collected measurement data includes \gls{rsrp} values, \gls{earfcn}, cell \gls{id} and \gls{gps} coordinates.
The \gls{earfcn} specifies the LTE band and the center frequency of the serving cell. To obtain these measurements, data were collected within each environment through both walking and driving surveys conducted on different days at different times, capturing variability in user mobility and traffic conditions representative of typical network usage. Due to accessibility constraints around each \gls{bs} and antenna, only driving-based surveys were conducted to collect \gls{rsrp} reference points around these locations.

To obtain cell-site information, we used the CellMapper application~\cite{cellmapperMobilityUnitedStates} and the Antenna Search database~\cite{AntennaSearchSearchCell}.  CellMapper, a crowd-sourced mobile application, provides cell-specific details such as cell types, IDs, uplink/downlink carrier frequencies, and approximate site locations. Using the cell IDs obtained from our measurements, we correlated them with the CellMapper database entries to locate the corresponding serving cells for our measured data. Because CellMapper provides only approximate locations, we refined this information by cross-referencing it with the Antenna Search database. This process enabled us to retrieve precise details, including exact GPS coordinates and tower heights, ensuring the accuracy of the \gls{bs} information.
\begin{figure*}[t!]
     \centering
     \begin{subfigure}[b]{0.29\textwidth} % Adjusted width
         \centering
         \includegraphics[width=\linewidth]{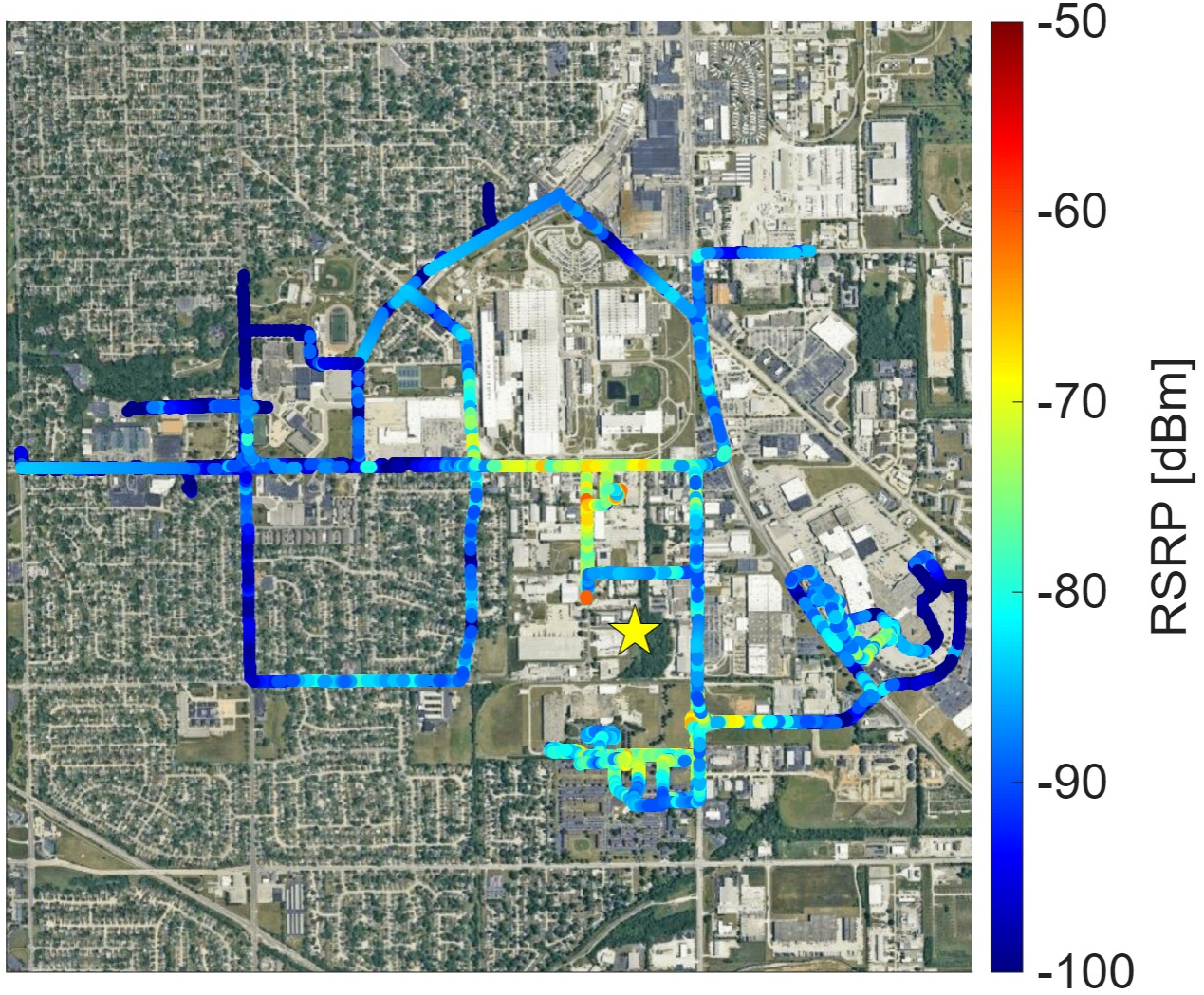}
        \vspace{-15pt}
         \caption{Measurements Points for BS1}
         \label{fig:relativersrpbs1}
     \end{subfigure}
     \hfill
     \begin{subfigure}[b]{0.29\textwidth} % Adjusted width
         \centering
         \includegraphics[width=\linewidth]{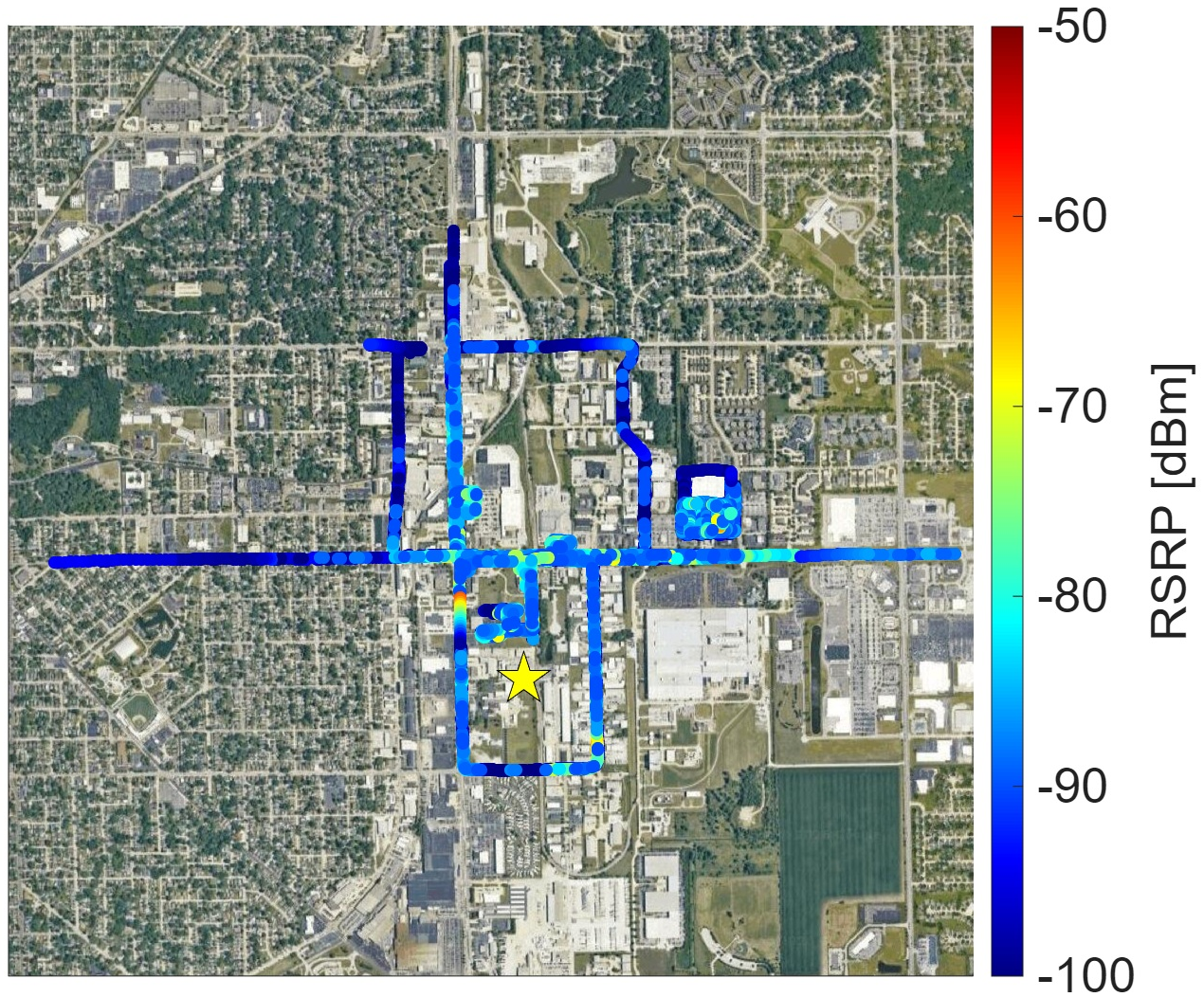}
        \vspace{-15pt}
         \caption{Measurements Points for BS2}
         \label{fig:relativersrpbs2}
     \end{subfigure}  
     \hfill
     \begin{subfigure}[b]{0.30\textwidth} % Adjusted width
         \centering
         \includegraphics[height=4.5cm]{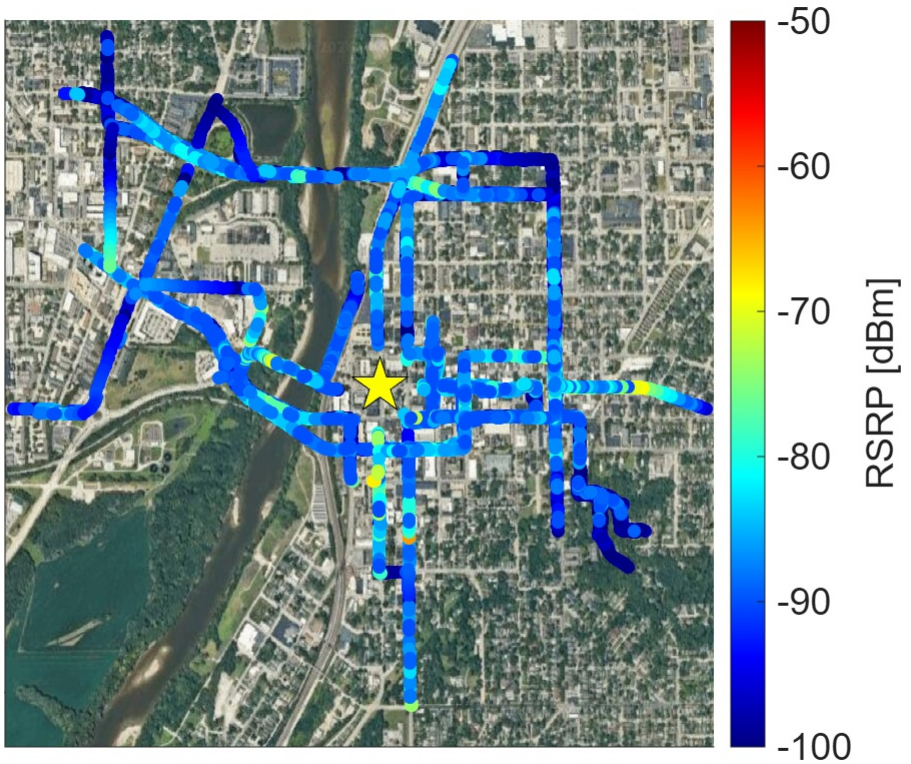}
        \vspace{-15pt}
         \caption{Measurements Points for BS3}
         \label{fig:relativersrpbs3}
     \end{subfigure}   
     % Caption and label
     \caption{\gls{rsrp} in \si{\dBm} for data points collected around three representative example BSs (out of 23 in our dataset). This figure illustrates the local measurement coverage from which LoS reference points are derived. Only the subset of points satisfying the filtering criteria—direct-path LoS, \gls{rsrp} $\ge$ $-80$ \si{\dBm}, and elevation angles within the main-lobe window—are retained as reference points. The yellow star indicates the BS location. Satellite imagery: Google Maps.}
     \label{fig:RSRPmap}
\end{figure*}
\subsection{$\Delta\text{\gls{rsrp}}$ Metric}
\label{subsec_processingrsrpdata}
The \gls{rsrp} in 4G LTE is a measure of the average power received in the resource elements (RE) that carry cell-specific reference signals (CRS).
The \gls{rsrp}, measured in \si{\dBm}, can be expressed as~\cite{zhangLargeScaleCellularCoverage2023,maengImpact3DAntenna2023}
\begin{equation} 
    \begin{split}
        \text{\gls{rsrp}}   
                =\text{-PL}+ \Delta \,,      
    \end{split}
\end{equation} 
where $PL$ represents the path loss in \SI{}{\deci\bel}, while $\Delta$ is an offset that captures the combined influence of site-specific parameters, such as transmit power, antenna gain, and antenna pattern as well as random fluctuations present in the propagation channel, as illustrated in Fig. \ref{fig:relativersrpbs}. %This offset accounts for variations in the signal caused by factors specific to the transmitter and receiver features.
Since transmitter and receiver parameters (e.g., transmit power, gain, antenna pattern, etc.) are unknowable for \gls{ue}-based \gls{rsrp} measurements, directly calculating path loss from \gls{rsrp} is not feasible, making it impossible to compare measured data with simulated path loss predictions. To address this challenge, we propose $\Delta \text{\gls{rsrp}}$ as a metric for data comparison. The use of  $\Delta \text{\gls{rsrp}}$ eliminates the dependency on unknown parameters and provides a consistent framework for comparing measurements across disparate datasets. Using $\Delta\text{\gls{rsrp}}$, all data points---whether produced from cell phone measurements or propagation models---are normalized to a common condition, so that both measurement-based \gls{rsrp} and model-based path loss can be expressed using the same target metric.

\subsection{Determining $\Delta \text{\gls{rsrp}}$ from \gls{rsrp} Measurement Data}

Accurately calculating $\Delta \text{\gls{rsrp}}$ from \gls{rsrp} measurements requires both (i) knowledge of the \gls{bs} antenna parameters---vertical 3 dB \gls{vbw}, horizontal pointing angle, gain, and downtilt---and (ii) \gls{rsrp} measurements taken under \gls{los} conditions within the antenna’s main lobe. Since the operator-specific antenna settings at each BS are proprietary, we infer effective, data-driven parameters from aggregated measurement data rather than attempting to recover exact antenna radiation characteristics. We initially select points with an unobstructed direct path to the BS, with \gls{rsrp} equal to or greater than \SI{-80}{\dBm}. These candidates are then used to infer an effective downtilt and dominant vertical beamwidth, which together define a stable reference subset. This subset is ultimately used to establish per-site baselines for $\Delta \text{\gls{rsrp}}$, as detailed below, enabling relative differencing that cancels site-dependent constants (e.g., TX gain and RX gain) while anchoring measurements to a stable main-lobe reference.

\begin{figure}[!h]
    \centering
\includegraphics[width=0.6\linewidth]{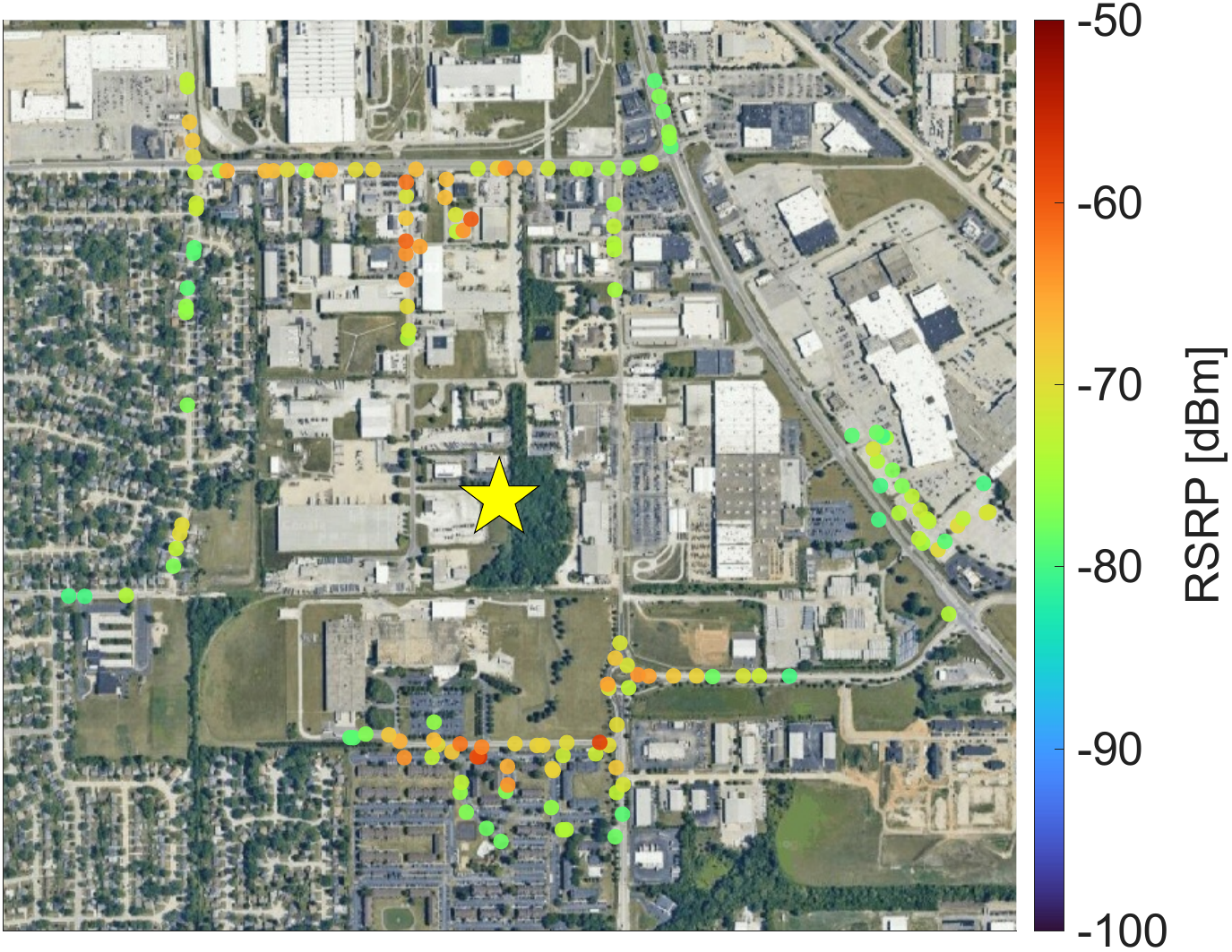}
    \caption{Retained measurement points around BS1, which uses a three-sector directional antenna, after applying all filtering criteria. This figure corresponds to the same site shown in ~\Cref{fig:relativersrpbs1} but displays only the subset of points used to form the $\Delta\text{\gls{rsrp}}$ reference. Satellite imagery: Google Maps.}
    \label{fig:filtered_points_bs1}
\end{figure}

\subsubsection{Identifying LoS Measurement Points}
We first select measurement points around each \gls{bs}—independently for each dataset and per sector/antenna—to form a set of \gls{los} reference points. For each point, we generate a lidar profile between the \gls{bs} and the receiver. A point is considered an \gls{los} candidate if the straight 3D path between the \gls{bs} and receiver is unobstructed in the lidar-derived terrain and clutter models. Additionally, measurements used for reference selection were collected in locations with visually unobstructed paths to the serving \gls{bs}, providing a practical consistency check.
Next, candidate points are required (i) to exhibit strong signal levels, with \gls{rsrp}~$\ge\SI{-80}{\dBm}$, and (ii) to fall within an effective main-lobe region, defined by the approximate \SI{3}{\dB} vertical beamwidth around the effective downtilt angle. We adopt $\SI{-80}{\dBm}$ as the threshold for   \gls{rsrp} in this study, consistent with prior measurement studies and industry signal-quality mappings~\cite{Chen2023, 3GPP36.133, 3GPP28.658}. While occasional fading may cause some \gls{los} samples to dip below this value, retaining only stronger points ensures reliability, and the averaging step described below mitigates any bias from these exclusions. Additionally, we compare each point’s geometric elevation angle against the vertical beamwidth window, excluding points outside the main lobe.
Only measurements meeting all criteria, namely lidar-consistent direct-path geometry, the \SI{-80} {\dBm} threshold, and main-lobe angular consistency, are retained as reference candidates for that site/sector.

\Cref{fig:RSRPmap} illustrates the local measurement coverage around three representative \gls{bs}s before filtering, showing how the raw measurement points extend around each site. For \gls{bs}1, the subset of retained points that satisfy all filtering criteria is shown in \Cref{fig:filtered_points_bs1}. Approximately 50 points per sector (about 150 in total for this BS) meet these conditions, primarily located between 400 m and 1 km from the site. From a simple geometric perspective, this distance range corresponds to elevation angles that are consistent with an effective antenna downtilt within typical deployment ranges ~\cite{3gpp36814,tekgul2021sample,tekgul2022uplink,athley2010impact,tmobile_agreement_mc2021,tmobile_agreement_orange2020,tmobile_usda_agreement_2018,verizon_agreement_mc2021,verizon_agreement_mc2023,jma_wireless,amphenol_antennas,commscope_antennas,cci_antennas,ericsson_antennas}. The retained points are first averaged within each sector to obtain sector-level \gls{los} references, and these sector averages are then averaged together to produce a single $\Delta\text{\gls{rsrp}}$ reference for the BS. Each sector-level average represents measurements collected over the local neighborhood of points within the main-lobe footprint, corresponding to a few meters of spatial separation between samples. This two-stage averaging helps mitigate small-scale fading, minor \gls{bs}/UE orientation uncertainties, and local terrain irregularities. The resulting averaged value constitutes the site-specific $\Delta\text{\gls{rsrp}}$ reference.

\subsubsection{BS Antenna Beamwidth and Downtilt}

Building on the \gls{los} candidate set identified for each \gls{bs}, we then account for antenna pattern effects. Because operator-specific antenna configurations are proprietary, we bound parameters using publicly available sources and estimate only the effective downtilt by minimizing the discrepancy between baseline channel model predictions and received power measurements. The downtilt search range is restricted to $0^\circ$–$15^\circ$, consistent with 3GPP specifications and operator/vendor studies~\cite{3gpp36814,tekgul2021sample,tekgul2022uplink,athley2010impact}. Public state–carrier agreements (T-Mobile, Verizon) provide antenna datasheets from JMA, Amphenol, CommScope, CCI, and Ericsson~\cite{tmobile_agreement_mc2021,tmobile_agreement_orange2020,tmobile_usda_agreement_2018,verizon_agreement_mc2021,verizon_agreement_mc2023,jma_wireless,amphenol_antennas,commscope_antennas,cci_antennas,ericsson_antennas}, which we use to bound 3~dB \gls{vbw} for the deployed antenna classes. All sites are three-sector deployments with omnidirectional horizontal coverage, so we focus on the vertical pattern and ignore horizontal beamwidth. To mitigate model bias, we consider five widely used baseline channel models and average the resulting downtilt estimates.

The primary transmission path (main lobe) is enforced via an elevation-angle filter. Let $\alpha(p)$ denote the geometric elevation angle from the \gls{bs} to a candidate \gls{los} point, $p$. For a candidate downtilt, $\theta$, we retain points satisfying
\begin{equation}
\alpha(p)\in\Big[\theta-\tfrac{\gls{vbw}}{2},\ \theta+\tfrac{\gls{vbw}}{2}\Big],
\end{equation}
and denote this main-lobe subset by $S_\theta$. Operationally, for each \gls{bs} (and, where applicable, each sector/antenna), we begin with all direct-path \gls{los} candidates in the dataset collected around that \gls{bs}/antenna and, for every candidate tilt $\theta$, filter by elevation angle to obtain $S_\theta$. Varying $\theta$, therefore, yields different subsets $S_\theta$, making explicit how the effective downtilt selects which \gls{los} points are aligned with the main lobe for comparison.

Although these points are geometrically \gls{los}, practical \gls{los} links in cellular deployments often deviate from ideal free-space due to multipath effects generated by near-ground clutter (foliage, vehicles, buildings), radiation-pattern/downtilt effects, and gentle terrain undulations. To capture these effects without over-reliance on any single model, we compare against five established path loss models (COST-eHata~\cite{itu2017sm2028}, \gls{sui}~\cite{abhayawardhana2005comparison}, 3GPP~\cite{3gpp36873}, \gls{ntia}-eHata~\cite{NTIA_eHata}, and \gls{spm}~\cite{forsk2018calibration}) during tilt selection. We conduct the comparison in a \emph{relative} (difference) form so site-dependent constants cancel. Specifically, for model $i$ and tilt $\theta$, the model-based relative prediction at $p\in S_\theta$ is
\begin{equation}
\Delta\text{\gls{rsrp}}_{\text{empirical},i}(\theta,p)
= -\,\text{PL}_i(p)\ +\ \overline{\text{PL}}_{i}(S_\theta),
\end{equation}
and the measured relative value is
\begin{equation}
\Delta\text{\gls{rsrp}}_{\text{real}}(p)
= \text{\gls{rsrp}}_{\text{meas}}(p)\ -\ \overline{\text{\gls{rsrp}}}_{\text{meas}}(S_\theta),
\end{equation}
where the $\overline{\text{\gls{rsrp}}}_{\text{meas}}(S_\theta)$ denotes the simple average over $S_\theta$. By differencing within the same main-lobe subset, unknown constants (such as RX gain and small orientation offsets) largely cancel, making the modeled and measured quantities directly comparable on a common relative scale without explicit knowledge of transmit power or antenna gains. Since the goal of this step is not to optimize the downtilt for any single path loss model, but rather to obtain a stable reference angle for subsequent relative \gls{rsrp} normalization, we avoid relying on any one model alone. Individual path loss models can exhibit different tilt-dependent behavior due to their distinct assumptions and calibration. Therefore, for each model $i$ and tilt $\theta$, we first compute the \gls{mae} between modeled and measured relative values on $S_\theta$, and then aggregate the model-specific errors using an equal-weight average across all five models,

\begin{equation}
\overline{\text{\gls{mae}}}(\theta) = \frac{1}{5} \sum_{i=1}^{5} \text{\gls{mae}}_{i}(\theta),
\end{equation}

to obtain a consensus, model-agnostic objective that reduces bias toward any single baseline model. The effective downtilt per \gls{bs} is finally selected as
\begin{equation}
\theta_{\text{est}}=\underset{\theta\in\{0^\circ,1^\circ,\dots,15^\circ\}}{\arg\min}\ \overline{\text{\gls{mae}}}(\theta),
\end{equation}

where $\theta_{\text{est}}$ should be interpreted as a consensus effective downtilt used to define a stable main-lobe-aligned reference subset, rather than as the optimum tilt under any one propagation model. This procedure standardizes the antenna configuration used in subsequent $\Delta\text{\gls{rsrp}}$ computations: \gls{vbw} is taken from datasheets, only $\theta$ is estimated, and the subset $S_\theta$ (and its means) is recomputed from the dataset collected around each \gls{bs}/antenna for every candidate tilt.

\subsubsection{$\Delta\text{\gls{rsrp}}$ Calculation}

After determining the effective downtilt $\theta_{\text{est}}$, we establish the per-\gls{bs} reference subset $S_{\theta_{\text{est}}}^{(b)}$ for each serving site $b$ based on the criteria outlined earlier: direct-path \gls{los} to the \gls{bs}, reliable measurements (\gls{rsrp}~$\ge\SI{-80}{\dBm}$), and geometric elevation angles within the main-lobe window $[\theta_{\text{opt}} \pm \gls{vbw}/2]$. From this subset, we calculate two baselines: (i) $\text{\gls{rsrp}}_{\text{REF}}(b)$, the average of measured \gls{rsrp} values within the subset; and (ii) for each empirical or synthetic model $i$, $\text{PL}_{\text{REF},i}(b)$, the average path loss of that model over the same subset:
\begin{equation}
\text{\gls{rsrp}}_{\text{REF}}(b)
= \frac{1}{\lvert S_{\theta_{\text{est}}}^{(b)}\rvert}
\sum_{q\in S_{\theta_{\text{est}}}^{(b)}}
\text{\gls{rsrp}}_{\text{meas}}(q).
\label{eq:rsrp-ref}
\end{equation}
\begin{equation}
\text{PL}_{\text{REF},i}(b)
= \frac{1}{\lvert S_{\theta_{\text{est}}}^{(b)}\rvert}
\sum_{q\in S_{\theta_{\text{est}}}^{(b)}}
\text{PL}_i(q).
\label{eq:pathloss-ref-i}
\end{equation}
After computing these per-site baselines, we compute $\Delta\text{\gls{rsrp}}$ for every sample in each of the six environments. Concretely, within environment $E_k$ and for each sample $p$ served by \gls{bs} $b$, we assign
\begin{equation}
\Delta\text{\gls{rsrp}}_{\text{real}}(p)
= \text{\gls{rsrp}}_{\text{meas}}(p)
- \text{\gls{rsrp}}_{\text{REF}}(b).
\label{eq:delta-real}
\end{equation}
\begin{equation}
\Delta\text{\gls{rsrp}}_{\text{empirical},i}(p)
= -\,\text{PL}_i(p)
+ \text{PL}_{\text{REF},i}(b).
\label{eq:delta-emp}
\end{equation}
At each real measurement location $p$, we also evaluate all empirical models to obtain $\text{PL}_i(p)$ and compute $\Delta\text{\gls{rsrp}}_{\text{empirical},i}(p)$ at the same point, yielding paired quantities $\big(\Delta\text{\gls{rsrp}}_{\text{real}}(p),\,\Delta\text{\gls{rsrp}}_{\text{empirical},i}(p)\big)$ referenced to the same baselines.
Both definitions use the same per-site reference subset $S_{\theta_{\text{opt}}}^{(b)}$ and subtract the subset mean of the corresponding quantity (\gls{rsrp} for measurements, PL for models). As a result, site-dependent constants that are approximately constant across points within the same main-lobe subset largely cancel each other out, providing a stable baseline. It also accommodates geometric variation within $S_{\theta_{\text{opt}}}^{(b)}$ (distance and elevation angle), so the reference reflects the local main-lobe geometry rather than a single point. Finally, because empirical/synthetic sources natively report path loss (dB) while field data report \gls{rsrp} (dBm), the relative forms in \eqref{eq:delta-real}–\eqref{eq:delta-emp} yield a single, unit-consistent target $\Delta\text{\gls{rsrp}}$ (dB) that is directly comparable across measurements, empirical models, and simulations and can be used uniformly for training and evaluation across all six environments.
\subsection{Creating Synthetic Datasets}
To construct the synthetic datasets, we employ a modeling framework using a large-scale path loss simulator~\cite{zhangLargeScaleCellularCoverage2023}. This simulator integrates high-resolution lidar datasets to capture detailed environmental and man-made features along all paths, enabling the prediction of large-scale fading. The simulator implements a physics-informed path loss model based on the \gls{ntia}-\gls{ehata} model~\cite{NTIA_eHata}, incorporating interactions between terrain profiles, man-made structures, and vegetation. 
The simulation area for each scenario was set to cover all measurement points and their surrounding environments. 
All carrier frequencies contained in the measurement dataset were included in the simulation.
\section{Model Creation}
This section describes the performance evaluation of several empirical models, as well as several \gls{ml} approaches using our measurement datasets.  Given the substantial influence of feature engineering on the predictive performance of \gls{ml} models, we define and extract geographical features known to influence path loss to ensure robust and accurate predictions.
\subsection{Feature Definition}
To extract geographic features, we utilized Indiana's statewide lidar datasets collected in 2018, along with the \gls{dsm} and \gls{dhm}, to capture detailed geographic profiles~\cite{PURR3707}. The \gls{dsm} provides elevation data above sea level, representing the vertical height of the tallest objects, such as buildings and trees, at each point. The \gls{dhm}, on the other hand, captures elevation above ground level, indicating the height of clutter, including structures and vegetation, at each point. Both the \gls{dsm} and \gls{dhm} datasets offer a spatial resolution of 5 feet.

\subsection{Feature Engineering}
As part of our feature engineering process, we identified and incorporated critical attributes that influence path loss predictions. 
Geographical attributes are defined to include both endpoint- and path-based characteristics, such as the relative height of the serving cell with respect to each reception point and the elevation angle. These engineered features, which combine radio and geographical information, are illustrated in \Cref{fig:relativersrpbs}. Each data sample corresponds to a unique \gls{bs}–measurement point  pair, and all features are generated in this per-point.
A detailed description of the features is provided below.
\begin{itemize}
    \item \textbf{Carrier Frequency:} The central frequency utilized for signal transmission. This feature is determined from the \gls{earfcn} attribute included in the measurement dataset.
    \item \textbf{Serving \gls{bs} Distance ($d_{\gls{bs}}$):} 
The horizontal distance measured between a measurement point and its associated serving \gls{bs}.

    \item \textbf{Relative \gls{bs} Height ($H_{BS}$):} 
    The elevation of the serving \gls{bs} relative to the measurement point.

    \item \textbf{Average Clutter Height ($H_C$):} Represents the average height of surrounding clutter relative to the measurement point. Neighboring points are defined as those located within a circular radius, R, centered on the measurement point. For our work, the radius is set to $R=\SI{50}{m}$. We use a \SI{50}{m} neighborhood around the measurement point to characterize local clutter, consistent with ranges specified in~\cite{itu2009radiowave} and appropriate for our predominantly flat environments.
    
    \item \textbf{Terrain Roughness:} 
    Characterizes the roughness of the terrain by calculating the difference between the $90\%$ and $10\%$ percentile ground elevation of the terrain profile. 
    \item \textbf{Transmitter Height Above Average Terrain (TxHAAT):} The elevation of the \gls{bs} relative to the average terrain elevation computed within a circular neighborhood of radius $R=\SI{50}{m}$ centered at the \gls{bs} location.
    
    \item \textbf{Ratio $\bm{\alpha}$}:
The ratio $\alpha$ is defined as the difference between the BS height and the height of the clutter around the measurement point, divided by the propagation path distance, given by $\frac{H^{'}_{bs} - H^{'}_{C}}{d_{BS}}$~\cite{reus-munsMachineLearningbasedMmWave2022}.

    \item \textbf{Ratio $\bm{\beta}$:}
     $\beta$ is the ratio of the average clutter height at the \gls{bs} relative to the propagation path distance, defined as $\frac{H_{BSc}}{d_{BS}}$~\cite{reus-munsMachineLearningbasedMmWave2022}.
    
    \item \textbf{Azimuth Angle of Arrival $\theta_{AoA}$:} The angle $\theta_{AoA}$ is given by $\tan^{-1}(\frac{x_{BS}-x_{UE}}{y_{BS}-y_{UE}})$, where $(x_{BS},y_{BS})$ and $(x_{UE},y_{UE})$ represent the coordinates of the \gls{bs} and the measurement point, respectively. 

    \item \textbf{Tilt Angle of Arrival $\phi_{AoA}$:} The angle $\phi_{AoA}$ is given by $\tan^{-1}(\frac{H^{'}_{ue} - H^{'}_{bs}}{d_{BS}})$ where $H^{'}_{bs}, H^{'}_{ue}$ are the elevations of \gls{bs} and the measurement point above sea level. 

    \item \textbf{First Diffraction Point $d_{\text{diff},1}$:} 
    This feature is the Euclidean distance from the BS to the first diffraction point along the direct ray path between the \gls{bs} and the measurement point. 
    \item \textbf{Last Diffraction Point $d_{\text{diff},\text{last}}$:} 
This feature is the Euclidean distance from the \gls{bs} to the last diffraction point along the direct ray path between the BS and the measurement point. 
    \item \textbf{Mean Terrain Height}:
This feature is the average height of the terrain along the propagation path between the \gls{bs} and the measurement point, computed from \gls{dsm} data by excluding clutter effects and referenced to the \gls{dsm}’s vertical datum. 

\item \textbf{Terrain Percentiles:} These features provide a statistical characterization of the terrain heights along the propagation path, highlighting variations in elevation that can impact signal propagation. The percentiles include:
    \begin{itemize}
        \item \textbf{25th Percentile:} The elevation value below which 25\% of the terrain elevations are distributed.
        \item \textbf{50th Percentile:} The median terrain height.
        \item \textbf{75th Percentile:} The elevation value below which 75\% of the terrain elevations are distributed.
    \end{itemize}

    \item \textbf{Blockage Percentage:}
This feature quantifies the fraction of the propagation path distance along the direct ray path between the \gls{bs} and the measurement point that are intersected by physical obstructions, such as buildings or vegetation. It is computed as the ratio of obstructed sample points to the total number of uniformly spaced sampled points along the path. 

\item \textbf{Diffraction Loss:} 
Quantifies the attenuation due to obstacles along the propagation path, calculated using the single- and multiple-knife-edge diffraction methods described in ITU-R P.526-13~\cite{ITU526}. 
\item \textbf{Propagation Condition (\gls{los}/\gls{nlos}):} A categorical feature indicating whether the propagation path between the \gls{bs} and the measurement point is \gls{los} or \gls{nlos}.
\end{itemize}

\begin{table*}[h]
\centering
\caption{\textsc{Geographic Regions, Sample Counts, and Descriptions of Measurement Datasets}}
\renewcommand{\arraystretch}{1.0}
\begin{tabular}{|c|c|c|c|c|}
\hline
\textbf{Area Name}          & \textbf{Label}     & \textbf{Samples} & \textbf{Description}                          & \textbf{Center Coordinates (Lat, Lon)} \\ \hline
\textbf{ACRE}               & Rural              & 71,919           & Rural farm area                               & 40.479450, -86.988450  \\ \hline
\textbf{Lindberg}           & Residential        & 128,038          & Residential area                              & 40.450450, -86.966000  \\ \hline
\textbf{Happy Hollow}       & Hilly 1            & 103,818          & Hilly suburban area with deciduous forest     & 40.458400, -86.894700  \\ \hline
\textbf{Southeast Lafayette}         & Industrial         & 104,981          & Industrial area                               & 40.421250, -86.853400  \\ \hline
\textbf{Downtown Lafayette} & Urban              & 41,838           & Urban area                                    & 40.419800, -86.893850  \\ \hline
\textbf{Valley}             & Hilly 2            & 85,586           & Hilly suburban area with medium-intensity development & 40.413300, -86.882150 \\ \hline
\end{tabular}
\label{tab:region_info}
\label{tab:dataset_summary}
\end{table*}

\subsection{Evaluation of Empirical and ML Models}
To evaluate the performance of both empirical and \gls{ml} propagation models, we compared their predictions against a large set of measured path loss data. The dataset consisted of 536,180 \gls{rsrp} measurements collected across diverse environments: rural, residential, industrial, two hilly environments, and urban areas, shown in Table~\ref{tab:dataset_summary}.  The diversity of scenarios provides a comprehensive basis for evaluating model performance under varying environmental conditions. For empirical models, we used the same five propagation models: the COST-eHata model~\cite{itu2017sm2028}, the \gls{sui} model~\cite{abhayawardhana2005comparison}, the 3GPP 3D channel model~\cite{3gpp36873}, the \gls{spm}~\cite{forsk2018calibration} and \gls{ntia}-eHata~\cite{NTIA_eHata}. For \gls{ml}-based approaches, we evaluated several models commonly used for tabular regression tasks in wireless communications and related fields. 

\textbf{Random Forest (RF)}~\cite{breiman2001random} is a bagging-based ensemble method that constructs multiple decision trees using bootstrap sampling and random feature selection at each split.

\textbf{Adaptive Boosting (AdaBoost)}~\cite{freund1995desicion} is a boosting algorithm that builds an ensemble in a sequential manner, with each learner focusing on correcting the mistakes of its predecessor.

\textbf{Huber Regression}~\cite{huber1992robust} is a robust regression method that minimizes Huber loss to achieve a balance between sensitivity to outliers and efficient parameter estimation. 

\textbf{Gradient Boosting Decision Trees (GBDT)}~\cite{friedman2001greedy} constructs additive models by training each tree to fit the residual errors of previous trees, enabling the model to capture complex patterns.

\textbf{Extreme Gradient Boosting (XGBoost)}~\cite{chen2016xgboost} is an optimized gradient boosting framework that incorporates regularization, parallelized tree construction, and sparsity-aware techniques for efficient large-scale learning.

\textbf{Light Gradient Boosting Machine (LightGBM)}~\cite{ke2017lightgbm} is a gradient boosting framework designed for high efficiency and low memory usage, making it particularly suitable for large-scale learning on structured datasets.

\textbf{Categorical Boosting (CatBoost)}~\cite{prokhorenkova2018catboost} is a gradient boosting method tailored for datasets with categorical features. It introduces strategies to reduce overfitting and improve prediction accuracy in such settings.

\textbf{Multi-Layer Perceptron (MLP)}~\cite{goodfellow2016deep} is a fully connected feedforward neural network capable of modeling non-linear relationships. 

The \gls{ml} models were trained only on the synthetic data but evaluated against 50\% of real-world measurements. This choice highlights the advantage of machine learning over empirical models by enabling scalable training with cost-effective synthetic data while maintaining strong performance on real-world measurements, without reliance on expensive data sources. The remaining half was set aside to maintain a consistent and balanced subset size across environments for fair comparisons, and to facilitate later experiments where real measurements are included in training to examine generalization performance.  The synthetic data was generated from our propagation simulator, which was configured with 8800 sampling points for Rural, 7500 points for Residential, 3900 points for Hilly 1, 5200 points for Industrial, 5800 points for Urban, and 8700 points for Hilly 2. Within each environment, simulations were conducted for all serving base stations and for the corresponding center frequencies observed in the measurements, which ranged from \SI{626.45}{\mega\hertz} to \SI{2525.8}{\mega\hertz}, to capture frequency-dependent propagation characteristics. After combining the results across environments, base stations, and frequencies, the final synthetic dataset contained approximately 5,059,074 data points.

For training ML models, we initially defined each transmitter–receiver propagation path as a training sample. For each path, we extracted the engineered features described above from the lidar data and other geographic information. These features were then paired with the corresponding $\Delta\text{\gls{rsrp}}$ value (measured or simulated), creating an input–output pair for supervised learning. This approach trained ML models to approximate the functional relationship between environmental  features and $\Delta\text{\gls{rsrp}}$, enabling them to understand not only the contribution of individual features, but also their combined effect on propagation behavior.
To evaluate the prediction performance of the \gls{ml} models, we use \gls{mae} and \gls{rmse}, which are expressed as
\begin{equation}
MAE = \frac{1}{n} \sum_{i=1}^{n} |L_i - \hat{L}_i|
\end{equation}
\begin{equation}
RMSE = \sqrt{\frac{1}{n} \sum_{i=1}^{n} (L_i - \hat{L}_i)^2}
\end{equation}
where $n$ is the number of \gls{rsrp} samples, $L_i$ is the true \gls{rsrp}, and $\hat{L}_i$ is the predicted \gls{rsrp}.

\begin{table*}[t]
    \centering
    \caption{\gls{mae} and \gls{rmse} comparison for empirical and machine learning models across different test environments. \gls{ml} models are trained on synthetic data generated to reflect the physical characteristics of the target environments without using real test measurements, highlighting generalization performance.} The best results in each column are \textbf{\underline{bold and underlined}}. Metrics are in decibels (dB).
    \renewcommand{\arraystretch}{0.75}
    \begin{tabular}{@{}l@{\hskip 10pt}lcccccccccccc@{}}
        \toprule
        & \multirow{3}{*}[-1.0em]{\textbf{Model}} & \multicolumn{12}{c}{\textbf{Metrics [dB] for Different Testing Datasets}} \\
        \cmidrule(lr){3-14}
        & & \multicolumn{2}{c}{\textbf{Rural}} & \multicolumn{2}{c}{\textbf{Hilly 1}} & \multicolumn{2}{c}{\textbf{Urban}} & \multicolumn{2}{c}{\textbf{Residential}} & \multicolumn{2}{c}{\textbf{Industrial}} & \multicolumn{2}{c}{\textbf{Hilly 2}} \\
        \cmidrule(lr){3-4} \cmidrule(lr){5-6} \cmidrule(lr){7-8} \cmidrule(lr){9-10} \cmidrule(lr){11-12} \cmidrule(lr){13-14}
        & & MAE & RMSE & MAE & RMSE & MAE & RMSE & MAE & RMSE & MAE & RMSE & MAE & RMSE \\
        \midrule
        \multirow{5}{*}{\textbf{Empirical}} 
        & COST-eHata        & 21.74 & 26.04 & 14.61 & 16.73 & 11.09 & 14.21 & 15.59 & 18.58 & 16.25 &19.71  & 12.39 & 14.49 \\
        & SUI              & 18.54 & 23.68 & 9.71 & 12.23 & 14.61 & 18.78 & 13.45 & 16.95 & 14.82 &18.70  & 10.13 & 13.04 \\
        & 3GPP             & 19.77 & 23.35 & 14.25 & 16.80 & 13.03 & 15.69 & 16.62 & 19.25 & 17.36 & 20.10 & 16.05 & 18.10 \\
        & SPM              & 17.68 & 22.05 & 9.83 &13.15  & 13.91 & 17.37 & 13.74 & 16.71 & 12.78 & 16.14 & 13.91 &  16.78\\
        & NTIA-eHata       & 18.84 & 22.82 & 9.01 & 12.40 & 11.40 & 14.03 & 13.24 & 16.89 & 16.77 & 20.80 & 9.76 & 12.65 \\
        \midrule
        \multirow{8}{*}{\textbf{ML}} 
        & AdaBoost         & 16.73 & 19.74 & 8.00 & 11.10 & 12.39 & 15.03 & 14.57 & 18.16 & 22.15 & 24.40 & 9.61 &  12.11\\
        & Random Forest    & 18.82 & 22.79 & 9.24 & 12.87 & 11.60 & 14.36 & 16.73 & 19.63 & 19.78 & 22.51 & 8.99 & 12.43 \\
        & Huber Regressor  & 10.85 & 13.45 & 7.37 & \textbf{\underline{10.34}} & 10.32 & 14.60 & 11.15 & 14.09 & 13.88 & 14.01 & 10.10 & 12.37 \\
        & Gradient Boosting & 12.51 & 15.02 & 7.66 & 11.00 & 8.46 & 10.58 & 11.97 & 15.01 &14.18  & 16.65 & 9.17 & 11.54 \\
        & XGBoost          & 10.73 & 13.58 & 7.12 & 10.49 & \textbf{\underline{7.39}} & 9.22 & 9.74 &\textbf{\underline{12.40}}  & 7.01 &  9.42 & \textbf{\underline{7.39}} & \textbf{\underline{9.24}} \\
        & LightGBM         & 9.85 & 12.51 & 7.12 & 10.50 & 7.42 & \textbf{\underline{9.21}} & 9.76 & 12.40 & 7.01 &  9.42 & 8.68 & 11.02 \\
        & CatBoost         & \textbf{\underline{9.81}} & \textbf{\underline{12.35}} & \textbf{\underline{7.10}} & 10.48 & 7.40 & 9.23 & \textbf{\underline{9.52}} & 12.43 & \textbf{\underline{7.00}} & \textbf{\underline{9.41}} & 8.57 & 10.83 \\
        & MLP              & 16.69 & 20.05 & 10.83 & 13.03 & 10.16 & 12.44 & 10.66 & 13.48 & 7.14 & 9.54 & 9.33 & 11.46 \\
        \bottomrule
    \end{tabular}

    \label{tab:mae_rmse_comparison}
\end{table*}

\subsection{Performance Comparison of Empirical Models and \gls{ml} Algorithms}

\begin{table*}[t]
    \centering
        \caption{Performance comparison across test environments using different training datasets. The best results in each column are \textbf{\underline{bold and underlined}}. Metrics are reported in dB.}
    \renewcommand{\arraystretch}{0.7}
    \begin{tabular}{@{}lcccccccccccc@{}}
        \toprule
        \textbf{Scenario} & \multicolumn{2}{c}{Rural} & \multicolumn{2}{c}{Hilly 1} & \multicolumn{2}{c}{Urban} & \multicolumn{2}{c}{Residential } & \multicolumn{2}{c}{Industrial} & \multicolumn{2}{c}{Hilly 2} \\
        \cmidrule(lr){2-3} \cmidrule(lr){4-5} \cmidrule(lr){6-7} \cmidrule(lr){8-9} \cmidrule(lr){10-11} \cmidrule(lr){12-13}
        & MAE & RMSE & MAE & RMSE & MAE & RMSE & MAE & RMSE & MAE & RMSE & MAE & RMSE \\
        \midrule
        % \rowcolor{lightergray} 
        SIM & 9.83 & 12.57  & 7.37 & 10.59 & 7.59 & 9.46 & 9.23 & 12.05 & 7.14 & 9.54 & 8.77 & 11.08 \\

        5\% Real  & 4.84 &6.29 & 4.48 &7.50 & 5.04 & 6.33& 5.28& 7.73&5.31 & 7.05&  3.03 &4.08 \\

        5\% Real + SMOTE & 4.67&6.00 &4.25 & 7.28 &4.96 & 6.17&5.23 & \textbf{\underline{7.54}}& 5.28&6.96 &2.98 &3.96 \\

        5\% Real + SIM & 4.80 &6.21&4.44 &7.46 & 4.98 &  6.15 &5.25 & 7.56& 5.26& \textbf{\underline{6.88}} & 3.00&4.02\\

        5\% Real + SMOTE + SIM & \textbf{\underline{4.64}} &\textbf{\underline{5.97}} & \textbf{\underline{4.24}} & \textbf{\underline{7.25}}&\textbf{\underline{4.95}} & \textbf{\underline{6.13}} &\textbf{\underline{5.21}}  & \textbf{\underline{7.54}}& \textbf{\underline{5.25}}& \textbf{\underline{6.88}}& \textbf{\underline{2.96}}&\textbf{\underline{3.94}} \\

        \bottomrule
    \end{tabular}

    \label{tab:performance_comparison}
\end{table*}

\newcommand{\colornum}[1]{%
  \pgfmathsetmacro{\minval}{2.53}%   Minimum value in table
  \pgfmathsetmacro{\maxval}{24.52}%  Maximum value in table
  \pgfmathsetmacro{\logmin}{ln(\minval)}% Natural log of min
  \pgfmathsetmacro{\logmax}{ln(\maxval)}% Natural log of max
  \pgfmathsetmacro{\logval}{ln(#1)}% Natural log of current value
  \pgfmathsetmacro{\percent}{50 - 50*((\logval - \logmin)/(\logmax - \logmin))}% Log-based gradient
  \pgfmathparse{int(round(\percent))}%
  \colorbox{green!\pgfmathresult}{#1}%
}

Table \ref{tab:mae_rmse_comparison} illustrates the performance of the empirical and \gls{ml} propagation models across the six measured environments. In general, we observe that the empirical models exhibit higher errors across varied site-specific environments due to their site-general nature. For example, while the SUI model achieves a moderate MAE of \SI{7.76}{\dB} in hilly open suburban environments, it performs poorly in other conditions, with MAE exceeding \SI{18}{\dB} in rural areas. Similarly, COST-eHata and 3GPP yield MAE above \SI{20}{\dB} in rural scenarios, illustrating their inflexibility in capturing terrain and clutter variations. \gls{ml}-based models achieve consistently lower errors by learning feature-based representations of propagation. This improved performance is enabled by training on large-scale synthetic datasets that densely cover the spatial domain of each environment, include all serving base stations and observed carrier frequencies, and span diverse propagation scenarios. Rather than attempting to reproduce the propagation model outputs, the \gls{ml} models learn a generalized mapping from rich physical features to relative \gls{rsrp}. When evaluated against real-world measurements, this learned representation can mitigate systematic biases in empirical models and provide more robust predictions.
While ensemble methods such as AdaBoost and Random Forest achieve competitive MAEs (ranging from about 10–14 dB), they can still exhibit errors exceeding \SI{17}{\dB} in more challenging environments, indicating potential overfitting or sensitivity to dataset biases. More advanced techniques like XGBoost, LightGBM, and CatBoost offer further improvements by effectively integrating diverse environmental features into their learning process, resulting in better adaptability across different geographical settings. By comparison, XGBoost, LightGBM, and CatBoost stand out for their strong predictive performance across multiple test environments, illustrating the robustness of gradient-boosted decision tree frameworks. For example, XGBoost achieves an MAE of approximately \SI{8.4}{\dB} in Urban and \SI{7.1}{\dB} in Industrial, while LightGBM demonstrates similarly strong performance with \SI{8.4}{\dB} in Rural and \SI{9.1}{\dB} in Residential, generally maintaining \gls{rmse} around 11–12 dB. However, CatBoost consistently delivers the best performance across all environments, with MAE values of \SI{7.07}{\dB} in Industrial, \SI{6.53}{\dB} in Hilly 2, \SI{8.36}{\dB} in Urban, and \SI{7.56}{\dB} in Hilly 1. These results highlight CatBoost’s ability to capture complex signal propagation patterns and generalize effectively across diverse scenarios. The performance is a result of CatBoost's efficient gradient boosting framework, which enable it to learn  signal propagation patterns from synthetic data and seamlessly transfer this knowledge to real-world conditions.

Given CatBoost’s superior generalizability and consistently low MAE and RMSE, we selected it as the primary \gls{ml} model for subsequent model development. The results in Table \ref{tab:mae_rmse_comparison} highlight an advantage of ML-based approaches over empirical models, showing that data-driven methods can achieve high accuracy without extensive data collection in the real world, as long as the synthetic datasets effectively capture the relevant propagation features.

\section{Proposed Framework and\\Generalization Evaluation}

\begin{algorithm}[ht]
\caption{Proposed Weighted Model Ensemble}
\small
\label{alg:ensemble}
\begin{algorithmic}[1]
\State \textbf{Input:}

\State $D_{\text{real}}$: Real measurements used in the experiment (5\% or 50\%, depending on setting)
\State $D_{\text{synth}}$: Synthetic data used for training
\State $D_{\text{real\_train}}, D_{\text{real\_val}} \subset D_{\text{real}}$
\State $D_{\text{synth\_train}}, D_{\text{synth\_val}} \subset D_{\text{synth}}$
\State $D_{\text{combined\_train}} = D_{\text{real\_train}} \cup D_{\text{synth\_train}}$
\State $D_{\text{val}} = D_{\text{real\_val}} \cup D_{\text{synth\_val}}$
\State $D_{\text{test}}$: Held-out real measurements (50\%)
\State \textbf{Split rule:}
\State For any dataset $D$ used for training, $D_{\text{train}} = 80\%$ and $D_{\text{val}} = 20\%$.
\State Initial weights: $w_1, w_2, w_3$ (equal-weight initialization)
\State Objective loss functions:
\State \hspace{1cm} MAE: $\text{MAE} = \frac{1}{n} \sum_{i=1}^{n} |y_{\text{true}, i} - y_{\text{ensemble}, i}|$
\State \hspace{1cm} RMSE: $\text{RMSE} = \sqrt{\frac{1}{n} \sum_{i=1}^{n} (y_{\text{true}, i} - y_{\text{ensemble}, i})^2}$
\State \textbf{Step 1: Train Individual Models (CatBoost)}
\State Train Model 1 on real data: $M_1 \gets \text{train}(D_{\text{real\_train}})$
\State Train Model 2 on synthetic data: $M_2 \gets \text{train}(D_{\text{synth\_train}})$
\State Train Model 3 on combined data: $M_3 \gets \text{train}(D_{\text{combined\_train}})$

\State \textbf{Step 2: Make Predictions on Validation Data}
\State Predict using Model 1 on validation data: $y_{\text{real\_val}} \gets M_1(D_{\text{real\_val}})$
\State Predict using Model 2 on validation data: $y_{\text{synth\_val}} \gets M_2(D_{\text{synth\_val}})$
\State Predict using Model 3 on validation data: $y_{\text{combined\_val}} \gets M_3(D_{\text{val}})$

\State \textbf{Step 3: Ensemble Predictions on Validation Data}
\( y_{\text{ensemble\_val}} = w_1 \cdot y_{\text{real\_val}} + w_2 \cdot y_{\text{synth\_val}} + w_3 \cdot y_{\text{combined\_val}} \).

\State \textbf{Step 4: Optimize Weights}
\State Optimize weights $w_1, w_2, w_3$ on validation data to minimize MAE or RMSE:
\State \hspace{1cm} $\min_{w_1, w_2, w_3} \text{Loss}(y_{\text{true\_val}}, y_{\text{ensemble\_val}})$

\State \textbf{Step 5: Evaluate Performance on Test Data}
\State Predict using Model 1 on test data: $y_{\text{real\_test}} \gets M_1(D_{\text{test}})$
\State Predict using Model 2 on test data: $y_{\text{synth\_test}} \gets M_2(D_{\text{test}})$
\State Predict using Model 3 on test data: $y_{\text{combined\_test}} \gets M_3(D_{\text{test}})$
\State Combine predictions on test data:
\( y_{\text{ensemble\_test}} = w_1 \cdot y_{\text{real\_test}} + w_2 \cdot y_{\text{synth\_test}} + w_3 \cdot y_{\text{combined\_test}} \).

\State Calculate performance metrics on test data:
\State \hspace{1cm} $\text{MAE}_{\text{final}} = \frac{1}{n} \sum_{i=1}^{n} |y_{\text{true\_test}, i} - y_{\text{ensemble\_test}, i}|$
\State \hspace{1cm} $\text{RMSE}_{\text{final}} = \sqrt{\frac{1}{n} \sum_{i=1}^{n} (y_{\text{true\_test}, i} - y_{\text{ensemble\_test}, i})^2}$

\State \textbf{Output:}
\State Optimized weights $w_1, w_2, w_3$
\State Final ensemble predictions $y_{\text{ensemble\_test}}$
\State Loss values: $\text{MAE}_{\text{final}}, \text{RMSE}_{\text{final}}$
\end{algorithmic}
\end{algorithm}
Generalization across diverse environments remains a critical challenge in wireless propagation modeling, particularly for \gls{ml} models, and especially when training data is limited. Models trained solely on measurement data often achieve high accuracy within their \emph{source} (measured) environment but struggle to maintain performance in the \emph{unseen} domains due to domain mismatch~\cite{vanleer2021improving,akrout2023domain}. On the other hand, using synthetic data for training provides a scalable alternative, enabling the generation of domain-specific datasets tailored to target environments.  To address these challenges, we propose an ensemble-based framework and testing processes, as well as the optimization of the ensemble weights.
The weighted ensemble model is as follows
\begin{equation}
\text{y}_{\text{ensemble}} = \text{w}_1 \cdot \text{y}_{\text{real}} + \text{w}_2 \cdot \text{y}_{\text{synthetic}} + \text{w}_3 \cdot \text{y}_{\text{combined}}
\end{equation}
where $w_1, w_2, w_3$ are the trainable weights, and $w_1 + w_2 + w_3 = 1$. The terms $y_{\text{real}}, y_{\text{synthetic}}, \text{and } y_{\text{combined}}$ represent the predictions made by the models trained on measurement data, synthetic data, and a combination of measurement and synthetic data, respectively. The proposed ensemble model creation procedure is described in Algorithm~\ref{alg:ensemble}.
Our motivation for this design arises from preliminary experiments~\cite{mohamed2024simulation} that indicate that directly mixing synthetic data from mismatched environments can sometimes degrade performance in a single model. The ensemble architecture instead exploits complementary strengths across models. Models trained on measurement data capture environment-specific propagation characteristics, while the model trained on synthetic data provides broad coverage patterns obtained through cost-effective generation. The ensemble model partially balances these advantages, offering a middle ground that balances domain-specific optimizations with generalizability to unseen environments. Through the trainable weighting mechanism, the ensemble learns how to optimally combine these perspectives, preserving the strong environment-specific accuracy of the measurement-based model while benefiting from the generalizability of the synthetic predictions. In most cases, the ensemble model provides an overall improvement in predictive performance by effectively combining information from real and synthetic data across different environments.
\begin{figure*}[htb]
    \centering
    % --- MAE Row ---
    \begin{subfigure}[b]{0.39\textwidth}
        \includegraphics[width=\linewidth]{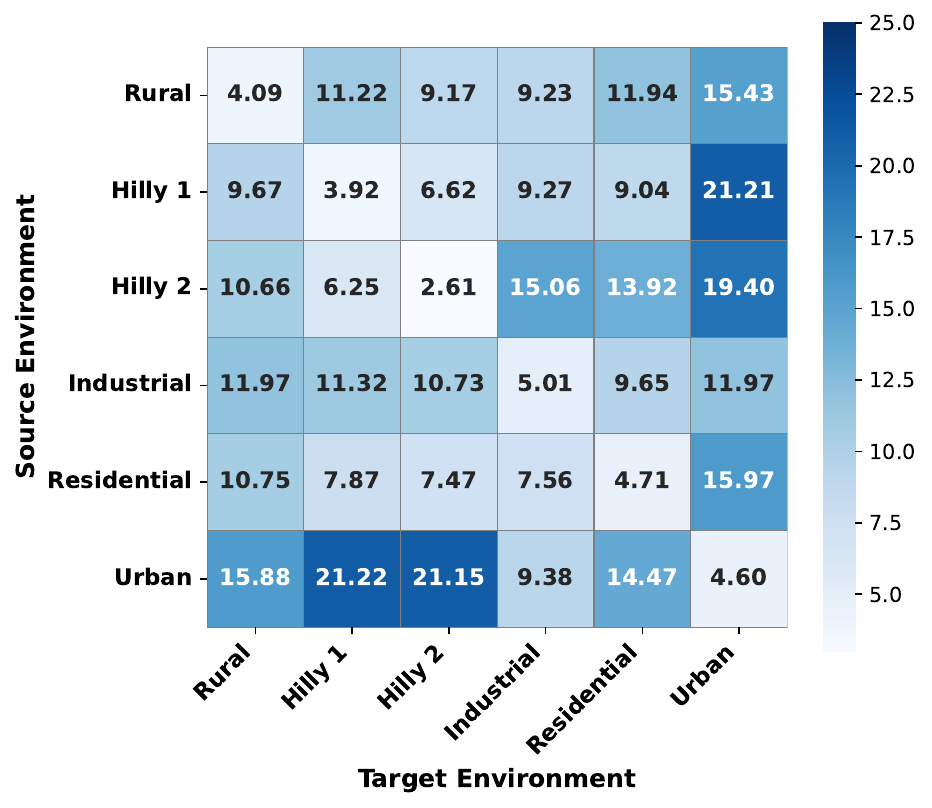}
        \caption{MAE, Source (R)}
        \label{fig:source_mae}
    \end{subfigure}
    \hfill
    \begin{subfigure}[b]{0.39\textwidth}
        \includegraphics[width=\linewidth]{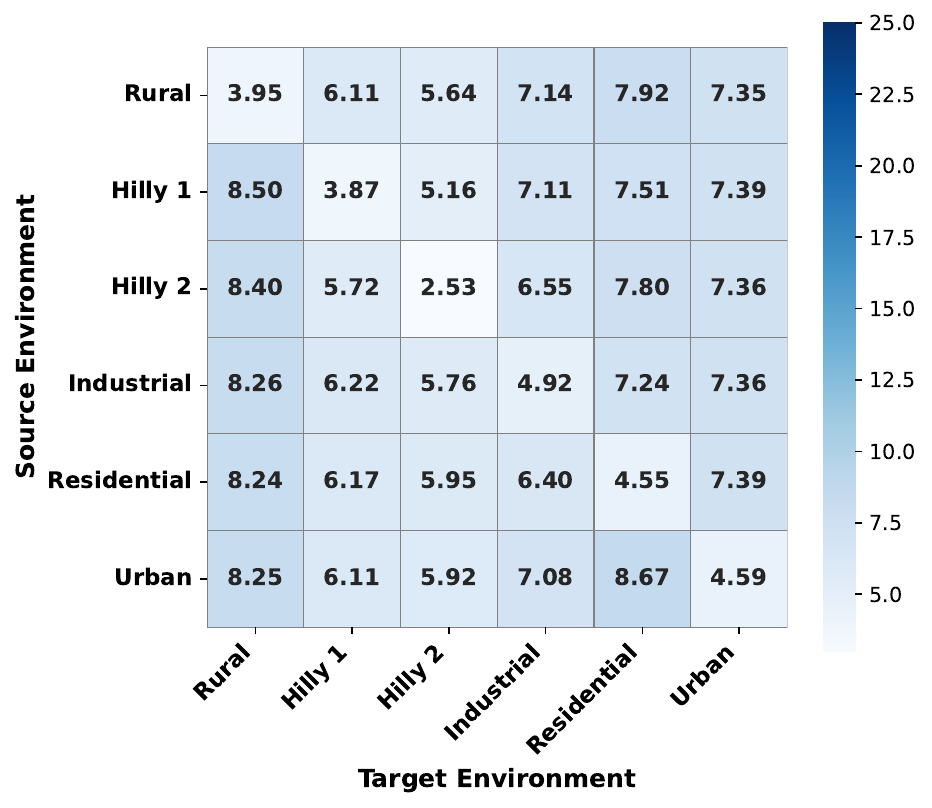}
        \caption{MAE, Source + Target (R+S)}
        \label{fig:combined_mae}
    \end{subfigure}
    % --- RMSE Row ---
    \begin{subfigure}[b]{0.39\textwidth}
        \includegraphics[width=\linewidth]{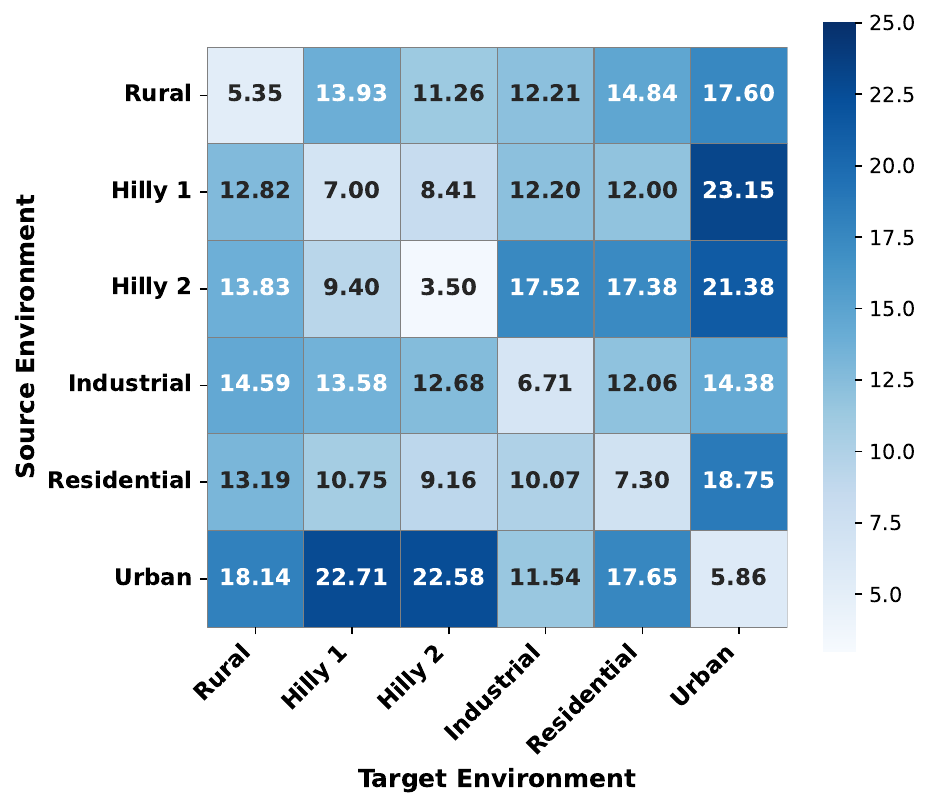}
        \caption{RMSE, Source (R)}
        \label{fig:source_rmse}
    \end{subfigure}
    \hfill
    \begin{subfigure}[b]{0.39\textwidth}
        \includegraphics[width=\linewidth]{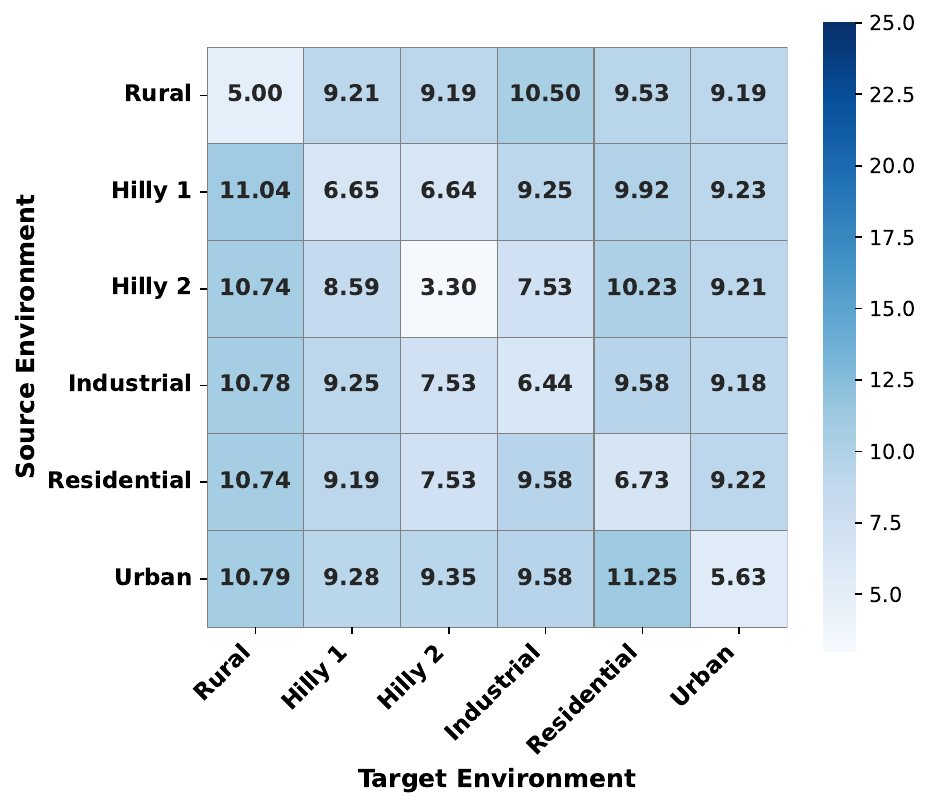}
        \caption{RMSE, Source + Target (R+S)}
        \label{fig:combined_rmse}
    \end{subfigure}
    \caption{
Error heatmaps of \gls{rsrp} prediction errors across environments (values in dB). Panels (A) and (C) correspond to models trained using real data only (Source~(R)),
and panels (B) and (D) correspond to models trained using real data augmented
with synthetic data generated for the target environment (Source~(R) + Target~(S)).
All results are evaluated on the same held-out 50\% of real measurement data.}
    \label{fig:confusion_matrices}
\end{figure*}
\subsection{Training Optimization with Limited Measurement Data}
A common problem when developing empirical models is \emph{data scarcity}, as the number of real measurement points will always be limited.  To evaluate the performance of our models under the condition of data scarcity, we limit the training set to only 5\% of the entire dataset. This 5\% is drawn randomly only from the portion of the data allocated for training, ensuring that the held-out 50\% test set remains untouched and consistent across experiments. The remaining 45\% of the data is not used in this scenario.  To counterbalance the limited measurement data, we evaluated another approach where the models were trained with the 5\% measurement data augmented with synthetic data generated via our simulation module (SIM). Finally, we evaluated another case where we used the \gls{smote}  to generate additional ``real-like" samples from the 5\% subset of available measurements. Rather than relying solely on purely synthetic data, our \gls{smote}-based approach uses linear interpolation in the feature space between neighboring samples to generate new points that reflect the distribution of real \gls{rsrp} measurements~\cite{douzas2018improving}.

The results given in Table \ref{tab:performance_comparison} demonstrate the improvements achieved by our simulation-aided data augmentation framework. A notable outcome is the improved prediction accuracy achieved even with training relies on only a small proportion of measured data. By integrating just 5\% of real-world measurement data into the training pipeline, our framework enhances the performance of models trained primarily on synthetic data. For example, the ``5\% Real" training dataset achieves better \gls{mae} and \gls{rmse} values across all test environments compared to the baseline ``SIM" training dataset. 

 The incorporation of \gls{smote} further amplifies these benefits by interpolating realistic samples and addressing imbalances in the data distribution. The ``5\% Real + \gls{smote}" training dataset outperforms models trained using real data alone in all tested scenarios. By bridging the gap between real and synthetic data, SMOTE allows the model to better capture the variability inherent in real-world environments, even when measurement data are sparse.

A key strength of our framework lies in its ability to integrate synthetic and measured data. Synthetic data is computationally efficient to generate but lacks the high-fidelity features and environmental characteristics captured in measured datasets. Measured data reflects real-world propagation effects with high spatial and temporal variability, yet remains limited in availability due to the constraints of data collection. By leveraging a balanced mix of both, as demonstrated in the ``5\% Real + SMOTE + SIM" training dataset, our approach achieves exemplary predictive accuracy across diverse scenarios.

Considering Table \ref{tab:performance_comparison}, in the Hilly 1 test environment, the \gls{mae} and \gls{rmse} values drop significantly from \SI{7.37}{\dB} and \SI{10.59}{\dB} (``SIM") to \SI{4.24}{\dB} and \SI{7.25}{\dB} (``5\% Real + SMOTE + SIM"). Similarly, in the Rural test environment, the ``SIM" configuration shows relatively high errors, with an \gls{mae} of \SI{9.83}{\dB} and an \gls{rmse} of \SI{12.57}{\dB}. By contrast, the ``5\% Real + SMOTE + SIM" approach significantly reduces these errors to \SI{4.64}{\dB} (\gls{mae}) and \SI{5.97}{\dB} (\gls{rmse}). 

Although synthetic data offers a scalable alternative to extensive measurement campaigns, it is not sufficient to fully represent the complexities of real-world environments. The ``SIM" training dataset consistently exhibits the highest error rates across all scenarios, underscoring its inability to fully capture these complexities. Our framework effectively addresses this limitation by supplementing synthetic data with limited real measurements and SMOTE-augmented samples, and significantly improves the model's generalizability.

\subsection{Cross-Environment Generalization Evaluation}
To evaluate cross-environment generalization, we analyze the performance of real-only and ensemble models across diverse propagation environments, as shown in Fig.~\ref{fig:confusion_matrices}. The main diagonal (from top-left to bottom-right) represents cases where both training and testing are conducted in the same environment, using 50\% of the real measurements for training and the remaining 50\% for testing. These cases produce the best results because the model is trained and tested on data from the same environment.  From Fig.~\ref{fig:confusion_matrices}a and \ref{fig:confusion_matrices}c, we observe that model performance degrades as the test environment becomes less similar to the training environment. When trained on Hilly~1 and evaluated on Hilly~2, the model achieves an \gls{mae} of \SI{6.62}{\dB} and an \gls{rmse} of \SI{8.41}{\dB}, indicating strong generalization under closely matched conditions. Testing the same model in the Urban environment yields substantially higher errors, with an \gls{mae} of \SI{21.21}{\dB} and an \gls{rmse} of \SI{23.15}{\dB}, highlighting the difficulty of generalizing across environments with fundamentally different propagation characteristics. These results emphasize the critical role of topographical and structural similarity in effective cross-environment generalization and highlight why real-only training can struggle under severe domain mismatch.

Fig.~\ref{fig:confusion_matrices}b and ~\ref{fig:confusion_matrices}d illustrate the advantages of the proposed ensemble approach, which augments real data with synthetic data tailored to the target environment. For example, training on Urban (R) and testing on Rural (R) results in an \gls{mae} of \SI{15.88}{\dB} and an \gls{rmse} of \SI{18.14}{\dB}, as illustrated in Fig.~\ref{fig:confusion_matrices}. As previously shown in Table~\ref{tab:mae_rmse_comparison}, training using Rural (S) reduces these errors to \SI{9.81}{\dB} and \SI{12.35}{\dB}, respectively. However, the ensemble approach, which augments available real data (R) with synthetic data (S), further reduces the \gls{mae} to \SI{8.25}{\dB} and the \gls{rmse} to \SI{10.79}{\dB},
 demonstrating a clear enhancement over the synthetic-only method. Similarly, when training using Urban (R) and testing on Hilly 1 (R), the real-only model (Urban (R)) results in \gls{mae} of \SI{21.22}{\dB} and \gls{rmse} of \SI{22.71}{\dB}. In contrast, the synthetic-only model (Hilly 1 (S)) achieves lower errors, with \gls{mae} of \SI{7.10}{\dB} and \gls{rmse} of \SI{10.48}{\dB}. The ensemble approach further improves the performance to \gls{mae} of \SI{6.11}{\dB} and \gls{rmse} of \SI{9.28}{\dB}, demonstrating its ability to leverage both real and synthetic data for improved performance effectively.

In the Residential environment, three scenarios highlight the relative performance of real and synthetic training data. First, Fig.~\ref{fig:confusion_matrices} illustrates training and testing on real Residential data provides in-environment baseline for the Residential case, yielding an \gls{mae} of \SI{4.71}{\dB} and an \gls{rmse} of \SI{7.30}{\dB}. This represents the best achievable performance when both training and evaluation are conducted within the same Residential environment. Second, Table~\ref{tab:mae_rmse_comparison} shows that training on Residential (S) and testing on Residential (R) results in an \gls{mae} of \SI{9.52}{\dB} and an \gls{rmse} of \SI{12.43}{\dB}. While less accurate than the fully real baseline, these results demonstrate that synthetic data provides a reasonable approximation of real-world performance and can serve as a practical substitute when direct measurements are unavailable or costly to collect. Third, training on real data from mismatched environments produces noticeably degraded accuracy when tested on Residential (R): for example, Fig.~\ref{fig:confusion_matrices} illustrates training with Urban (R) leads to an \gls{mae} of \SI{14.47}{\dB} and an \gls{rmse} of \SI{17.65}{\dB}, while Hilly~2 (R) yields an \gls{mae} of \SI{13.92}{\dB} and an \gls{rmse} of \SI{17.38}{\dB}. Together, these results show that environment-aligned synthetic data generalizes more effectively to the Residential case than real data collected in mismatched environments.

The ensemble model mitigates these cross-domain limitations. When combining Urban (R) with Residential (S), the ensemble reduces the error to an \gls{mae} of \SI{8.67}{\dB} and an \gls{rmse} of \SI{11.25}{\dB}. Further improvement is achieved when the ensemble incorporates Hilly 2 (R), reducing the \gls{mae} to \SI{7.80}{\dB} and the \gls{rmse} to \SI{10.23}{\dB}. These results demonstrate the ensemble's capability to effectively integrate real and synthetic data, outperforming mismatched real and synthetic-only models. In challenging cross-environment scenarios, the ensemble model consistently achieves lower \glspl{mae} and \glspl{rmse}, which demonstrates its adaptability and generalizability. These cross-environment experiments demonstrate that augmenting real data with environment-aligned synthetic data yields up to a 50\% reduction in \gls{mae} relative to real-only training and up to a 25\% reduction relative to synthetic-only training, particularly in challenging domain-mismatched scenarios.

The strength of the ensemble lies in its ability to combine the complementary advantages of both data types. Measurement data captures environment-specific details, such as multipath effects, clutter characteristics, and terrain features, all of which are crucial for accurate prediction within source environments. Synthetic data provides broader generalization across diverse domains and supplies valuable supplemental samples for underrepresented or previously unseen areas.
The ensemble adapts to a target environment by employing a dynamic weighting scheme. It prioritizes the real model when the source and target domains are closely matched, and shifts toward the synthetic model when domain differences increase. This adaptive fusion preserves high source-domain accuracy and improves generalization across mismatched terrains.
\section{Conclusion}
This study has demonstrated the effectiveness of an ensemble-based machine learning approach for path loss prediction, integrating measurement and synthetic data to address the challenges of data scarcity and environmental variability. Using a combination of measurements, simulation-generated data, and advanced data augmentation techniques, we have demonstrated an improved, robust, and generalizable path loss prediction model.  The use of lidar data to derive relevant features has proven essential to capture the physical and geographical details that influence the propagation of wireless signals.  The model produces consistent performance improvements over purely \gls{ml} or empirical models in diverse environments, including urban, rural, industrial, and residential scenarios. 
 By advancing the capabilities of augmented machine learning in the domain of wireless communication, this research contributes to the development of more reliable and efficient network planning tools, ultimately supporting the deployment of next-generation wireless networks.
\bibliographystyle{IEEEtranDOI} % This should fix the "DOI not showing up" issue. Note that you will need the doi field in the BibTex file. 
\bibliography{IEEEabrv,references}

@STRING{IEEE_J_WCOM       = "{IEEE} Trans. Wireless Commun."}

@STRING{IEEE_J_MC         = "{IEEE} Trans. Mobile Comput."}

@STRING{IEEE_J_AP         = "{IEEE} Trans. Antennas Propag."}

@STRING{IEEE_J_OAP         = "IEEE Open J. Antennas Propag."}

@STRING{IEEE_M_COM        = "{IEEE} Commun. Mag."}

@STRING{IEEE_O_CSTO       = "{IEEE} Commun. Surveys Tuts."}

@IEEEtranBSTCTL{BSTcontrol,
	CTLuse_forced_etal       = "yes",
	CTLmax_names_forced_etal = "6",
	CTLnames_show_etal       = "1",
        CTLuse_alt_dash = "no"
}

@ARTICLE{ghosh2025,
  author={Doğan-Tusha, Seda and Tusha, Armed and Rochman, Muhammad Iqbal and Nasiri, Hossein and Ghosh, Monisha},
  journal={IEEE Communications Magazine}, 
  title={{Spectrum Sharing Characterization Using Smartphones: Exploring 6 GHz Sharing Through Large-Scale Wi-Fi 6E Measurements}}, 
  year={2025},
  volume={63},
  number={2},
  pages={70-76},
  doi={10.1109/MCOM.001.2400325}}

@article{Chen2023,
  author    = {Chen, Yang and Wei, Yuxin and Chen, Ming and Shi, Jinhong and Shi, Shaofu and Liu, Bin and Li, Jianhua and Wang, Mingfeng},
  title     = {{Measurement and Analysis of 4G/5G Mobile Signal Coverage in a Metropolitan Area}},
  journal   = {Sensors},
  volume    = {23},
  number    = {15},
  pages     = {6594},
  year      = {2023},
  doi       = {10.3390/s23156594}
}

@techreport{3GPP36.133,
  author      = {{3GPP}},
  title       = {{E-UTRA; Requirements for Support of Radio Resource Management (Release 18)}},
  number      = {TS~36.133},
  year        = {2023},
  month       = jan,
  note        = {v18.0.0},
  institution = {3rd Generation Partnership Project}
}

@techreport{3GPP28.658,
  author       = {{3GPP}},
  title        = {{Telecommunication management; LTE Radio Access Network (RAN) Performance Measurements}},
  number       = {TS 28.658},
  version      = {16.3.0},
  year         = {2021},
}

@book{itu2009radiowave,
  author    = {{International Telecommunication Union}},
  title     = {Radiowave Propagation Information for Designing Terrestrial Point-to-Point Links},
  publisher = {International Telecommunication Union},
  address   = {Geneva, Switzerland},
  year      = {2008},
  note      = {ITU-R Handbook, Edition 2008}
}

@Article{jaeckelQuaDRiGa3DMultiCell2014,
  author    = {Jaeckel, Stephan and Raschkowski, Leszek and B{\"o}rner, Kai and Thiele, Lars},
  journal   = IEEE_J_AP,
  title     = {{QuaDRiGa}: A {3-D} multi-cell channel model with time evolution for enabling virtual field trials},
  year      = {2014},
  number    = {6},
  pages     = {3242--3256},
  volume    = {62},
  publisher = {IEEE},
}

@article{phillipsSurveyWirelessPath2013,
  title={A survey of wireless path loss prediction and coverage mapping methods},
  author={Phillips, Caleb and Sicker, Douglas and Grunwald, Dirk},
  journal=IEEE_O_CSTO,
  volume={15},
  number={1},
  pages={255--270},
  year={2012},
  publisher={IEEE}
}

@Article{zhangLargeScaleCellularCoverage2023,
  author    = {Zhang, Yaguang and Krogmeier, James V and Anderson, Christopher R and Love, David J},
  journal   = IEEE_J_WCOM,
  title     = {Large-scale cellular coverage simulation and analyses for follow-me {UAV} data relay},
  year      = {2023},
  publisher = {IEEE},
}

@Article{maengImpact3DAntenna2023,
  author    = {Maeng, Sung Joon and Kwon, Hyeokjun and Ozdemir, Ozgur and G{\"u}ven{\c{c}, Ismail}},
  journal   = IEEE_J_OAP,
  title     = {Impact of {3D} antenna radiation pattern in {UAV} air-to-ground path loss modeling and {RSRP}-based localization in rural area},
  year      = {2023},
  publisher = {IEEE},
}

@article{zhangChallengesOpportunitiesFuture2021a,
  title={Challenges and opportunities of future rural wireless communications},
  author={Zhang, Yaguang and Love, David J and Krogmeier, James V and Anderson, Christopher R and Heath, Robert W and Buckmaster, Dennis R},
  journal=IEEE_M_COM,
  volume={59},
  number={12},
  pages={16--22},
  year={2021},
  publisher={IEEE}
}

@article{zhang2022artificial,
  title={An artificial intelligence radio propagation model based on geographical information},
  author={Zhang, Hua and Dong, Jiangbo and Liu, Xingxu and Liu, Jianfei and Zhang, Xincheng},
  journal=IEEE_J_AP,
  volume={70},
  number={12},
  pages={12049--12060},
  year={2022},
  publisher={IEEE}
}

@Article{masood2022interpretable,
  author    = {Masood, Usama and Farooq, Hasan and Imran, Ali and Abu-Dayya, Adnan},
  journal   = IEEE_J_MC,
  title     = {Interpretable {AI}-based large-scale {3D} pathloss prediction model for enabling emerging self-driving networks},
  year      = {2022},
  number    = {7},
  pages     = {3967--3984},
  volume    = {22},
  publisher = {IEEE},
}

@Article{chawla2002smote,
  author  = {Chawla, Nitesh V. and Bowyer, Kevin W. and Hall, Lawrence O. and Kegelmeyer, W. Philip},
  journal = {J. Artificial Intelligence Research},
  title   = {{SMOTE}: synthetic minority over-sampling technique},
  year    = {2002},
  pages   = {321--357},
  volume  = {16},
}

@article{breiman2001random,
  title={Random forests},
  author={Breiman, Leo},
  journal={Machine learning},
  volume={45},
  pages={5--32},
  year={2001},
  publisher={Springer}
}

@article{friedman2001greedy,
  title={Greedy function approximation: a gradient boosting machine},
  author={Friedman, Jerome H},
  journal={Annals of Statistics},
  pages={1189--1232},
  year={2001},
  publisher={JSTOR}
}

@article{ke2017lightgbm,
  title={{LightGBM: A highly efficient gradient boosting decision tree}},
  author={Ke, Guolin and Meng, Qi and Finley, Thomas and Wang, Taifeng and Chen, Wei and Ma, Weidong and Ye, Qiwei and Liu, Tie-Yan},
  journal={Advances in Neural Inf. Process. Syst.},
  volume={30},
  year={2017}
}

@article{prokhorenkova2018catboost,
  title={{CatBoost: unbiased boosting with categorical features}},
  author={Prokhorenkova, Liudmila and Gusev, Gleb and Vorobev, Aleksandr and Dorogush, Anna Veronika and Gulin, Andrey},
  journal={Advances in Neural Inf. Process. Syst.},
  volume={31},
  year={2018}
}

@misc{COST231,
  author       = {{COST 231}},
  title        = {Urban Transmission Loss Models for Mobile Radio in the 900 and {1800 MHz} Band},
  howpublished = {COST 231 TD (90), 119 Rev.~2, The Hague, The Netherlands},
  year         = {1991}
}

@incollection{huber1992robust,
  author    = {Huber, Peter J.},
  title     = {Robust Estimation of a Location Parameter},
  booktitle = {Breakthroughs in Statistics: Methodology and Distribution},
  editor    = {Kotz, Samuel and Johnson, Norman L.},
  publisher = {Springer},
  address   = {New York, NY},
  year      = {1992},
  pages     = {492--518},
  doi       = {10.1007/978-1-4612-4380-9_35}
}

@inproceedings{reus-munsMachineLearningbasedMmWave2022,
  title={Machine learning-based {mmWave} path loss prediction for urban/suburban macro sites},
  author={Reus-Muns, Guillem and Du, Jinfeng and Chizhik, Dmitry and Valenzuela, Reinaldo and Chowdhury, Kaushik R},
  booktitle={GLOBECOM 2022-2022 IEEE Global Communications Conference},
  pages={1429--1434},
  year={2022},
  organization={IEEE}
}

@InProceedings{zhangSimulationAidedMeasurementBasedChannel2020,
  author       = {Zhang, Yaguang and Tan, John A. and Dorbert, Bryan M. and Anderson, Christopher R. and Krogmeier, James V.},
  booktitle    = {GLOBECOM 2020-2020 IEEE Global Commun. Conf.},
  title        = {Simulation-aided measurement-based channel modeling for propagation at 28 {GHz} in a coniferous forest},
  year         = {2020},
  organization = {IEEE},
  pages        = {1--6},
}

@InProceedings{abhayawardhana2005comparison,
  author       = {Abhayawardhana, V. S. and Wassell, I. J. and Crosby, D. and Sellars, M. P. and Brown, M. G.},
  booktitle    = {2005 IEEE 61st Veh. Technol. Conf.},
  title        = {Comparison of empirical propagation path loss models for fixed wireless access systems},
  year         = {2005},
  organization = {IEEE},
  pages        = {73--77},
  volume       = {1},
}

@InProceedings{thrane2020deep,
  author       = {Thrane, Jakob and Sliwa, Benjamin and Wietfeld, Christian and Christiansen, Henrik L.},
  booktitle    = {GLOBECOM 2020-2020 IEEE Global Commun. Conf.},
  title        = {Deep learning-based signal strength prediction using geographical images and expert knowledge},
  year         = {2020},
  organization = {IEEE},
  pages        = {1--6},
}

@InProceedings{mohamed2024simulation,
  author       = {Mohamed, Ahmed P. and Lee, Byunghyun and Zhang, Yaguang and Hollingsworth, Max and Anderson, C. Robert and Krogmeier, James V. and Love, David J.},
  booktitle    = {ICC 2024-IEEE Intl. Conf. Communications},
  title        = {Simulation-enhanced data augmentation for machine learning pathloss prediction},
  year         = {2024},
  organization = {IEEE},
  pages        = {4863-4868},
}

@misc{dietterich2000Ensemble,
  author       = {Dietterich, T. G.},
  howpublished = {Ensemble Methods in Machine Learning. In: Multiple Classifier Systems. MCS 2000. Lecture Notes in Computer Science, vol 1857. Springer, Berlin, Heidelberg, 2000,  \url{https://doi.org/10.1007/3-540-45014-9_1}}
}

@article{rokach2010ensemble,
  title={Ensemble-based classifiers},
  author={Rokach, Lior},
  journal={Artificial intelligence review},
  volume={33},
  pages={1--39},
  year={2010},
  publisher={Springer}
}

@InProceedings{tekgul2021sample,
  author       = {Tekgul, Ezgi and Novlan, Thomas and Akoum, Salam and Andrews, Jeffrey G.},
  booktitle    = {2021 IEEE Inf. Theory Workshop (ITW)},
  title        = {Sample-efficient learning of cellular antenna parameter settings},
  year         = {2021},
  organization = {IEEE},
  pages        = {1--6},
}

@InProceedings{tekgul2022uplink,
  author       = {Tekgul, Ezgi and Novlan, Thomas and Akoum, Salam and Andrews, Jeffrey G.},
  booktitle    = {IEEE Global Commun. Conf.},
  title        = {Uplink-downlink joint antenna optimization in cellular systems with sample-efficient learning},
  year         = {2022},
  pages        = {6499--6504},
}

@InProceedings{athley2010impact,
  author       = {Athley, Fredrik and Johansson, Martin N.},
  booktitle    = {2010 IEEE 71st Veh. Tech. Conf.},
  title        = {Impact of electrical and mechanical antenna tilt on {LTE} downlink system performance},
  year         = {2010},
  organization = {IEEE},
  pages        = {1--5},
}

@article{freund1995desicion,
  title={A decision-theoretic generalization of on-line learning and an application to boosting},
  author={Freund, Yoav and Schapire, Robert E},
  journal={Journal of computer and system sciences},
  volume={55},
  number={1},
  pages={119--139},
  year={1997},
  publisher={Elsevier}
}

@inproceedings{chen2016xgboost,
  title={{XGBoost: A scalable tree boosting system}},
  author={Chen, Tianqi and Guestrin, Carlos},
  booktitle={Proceedings of the 22nd ACM SIGKDD Intl. Conf. Knowledge Discovery and Data Mining},
  pages={785--794},
  year={2016}
}

@InProceedings{hollingsworth2024repurposing,
  author       = {Hollingsworth, Max and Zhang, Yaguang and Schumann, Todd and Anderson, Christopher R. and Cotton, Michael and Kim, Seyeon and Ha, Sangtae and Grunwald, Dirk},
  booktitle    = {2024 IEEE Intl. Symposium on Dynamic Spectrum Access Networks (DySPAN)},
  title        = {Repurposing cellular reference signals: accurate {RSRP} measurements with mobile phones},
  year         = {2024},
  organization = {IEEE},
  pages        = {45--50},
  doi          = {10.1109/DySPAN60163.2024.10632813}

}

@techreport{ITU526,
  author      = {{International Telecommunication Union}},
  title       = {Propagation by diffraction},
  institution = {International Telecommunication Union},
  type        = {Recommendation ITU-R},
  number      = {P.526-16},
  month       = nov,
  year        = {2025}
}

@manual{forsk2018calibration,
  author       = {Forsk},
  title        = {Atoll 3.3.2 User Manual for Radio Networks},
  year         = {2016},
  address      = {France},
  note         = {AT332\_UMR\_E0. [Online]. Available: \url{https://github.com/aprincemohamed/DeepLearningBasedCellularCoverageMap/raw/main/Atoll-3-3-2-User-Manual-Radio.pdf}}
}

@misc{AntennaSearchSearchCell,
  author       = {{AntennaSearch}},
  title        = {Search for Cell Towers \& Antennas},
  howpublished = {\url{https://www.antennasearch.com}},
  year         = {2025}
}

@misc{cellmapperMobilityUnitedStates,
  author       = {{CellMapper.net}},
  title        = {{AT\&T Mobility (United States of America) -- Cellular Coverage and Tower Map}},
  howpublished = {\url{https://www.cellmapper.net/map?MCC=-1&MNC=1}},
  year         = {2025},
}

@misc {PURR3707,
	title = {Indiana Statewide Normalized Digital Height Model (2016-2019)},
	month = {Feb},
	url = {https://purr.purdue.edu/publications/3708/1},
	year = {2021},
	doi = {doi:/10.4231/QAA5-6J29},
	author = {Jinha Jung and Sungchan Oh }
}

@Misc{NTIA_eHata,
  author       = {{National Telecommunications and Information Administration, U.S. Department of Commerce}},
  howpublished = {\url{https://github.com/NTIA/ehata.git}},
  title        = {The Extended {Hata (eHata)} Urban Propagation Model},
}

@misc{tmobile_agreement_mc2021,
  author = {{Montgomery County Transmission Facilities Coordination Group (TFCG)}},
  title  = {{T-Mobile} Minor Modification Site Application {MC2021071498}},
  howpublished = {{Montgomery County, Maryland}},
  year   = {2021},
  url    = {https://montgomerycountytfcg.s3.amazonaws.com/Applications/MC2021071498+Application.pdf}
}

@Misc{tmobile_agreement_orange2020,
  author = {{Connecticut Siting Council}},
  title  = {{T-Mobile} application: site filing for orange center road},
  year   = {2020},
  url    = {https://portal.ct.gov/-/media/csc/2_ems-medialibrary/orange/orangecenterrd/t-mobile/em-t-mobile-107-200817_filing_525-orange-ctr-rd-ct11412a.pdf},
}

@Misc{verizon_agreement_mc2023,
  author = {{Montgomery County Transmission Facilities Coordination Group (TFCG)}},
  title  = {{Verizon Wireless} colocation application: Site application {MC2023102222},
            {Montgomery County, Maryland}},
  year   = {2023},
  url    = {https://montgomerycountytfcg.s3.amazonaws.com/Applications/MC2023102222+Application.pdf},
}

@Misc{verizon_agreement_mc2021,
  author = {{Montgomery County Transmission Facilities Coordination Group (TFCG)}},
  title  = {{Verizon Wireless} minor modification application: Site MC2021111599,
            {Montgomery County, Maryland}},
  year   = {2021},
  url    = {https://montgomerycountytfcg.s3.amazonaws.com/Applications/MC2021111599+Application.pdf},
}

@Misc{tmobile_usda_agreement_2018,
  author = {{National Capital Planning Commission (NCPC)}},
  title  = {{Beltsville Agricultural Research Center}: replacement of three antennas at the {T-mobile} telecommunications facility},
  year   = {2018},
  url    = {https://www.ncpc.gov/files/projects/2018/7979_Beltsville_Agricultural_Research_Center_Replacement_of_Three_Antennas_at_the_T-Mobile_Telecommunications_Facility_-_Water_Tower_286_Submission_Materials_Jul2018.pdf},
}

@Misc{jma_wireless,
  author = {{JMA Wireless}},
  note   = {Accessed: January 2, 2025},
  title  = {{JMA Wireless} antenna solutions},
  year   = {2025},
  url    = {https://www.jmawireless.com/},
}

@misc{amphenol_antennas,
  author       = {Amphenol Antenna Solutions},
  title        = {Antenna products and specifications},
  year         = {2025},
  note         = {Accessed: January 2, 2025},
  url          = {https://amphenol-antennas.com/}
}

@Misc{commscope_antennas,
  author = {CommScope},
  note   = {Accessed: January 2, 2025},
  title  = {{CommScope} antenna portfolio},
  year   = {2025},
  url    = {https://www.commscope.com/},
}

@Misc{cci_antennas,
  author = {{Communication Components Inc. (CCI)}},
  note   = {Accessed: January 2, 2025},
  title  = {{CCI} antenna solutions},
  year   = {2025},
  url    = {https://www.cciproducts.com/},
}

@misc{ericsson_antennas,
  author       = {Ericsson},
  title        = {Ericsson antenna systems},
  year         = {2025},
  note         = {Accessed: January 2, 2025},
  url          = {https://www.ericsson.com/}
}

@book{goodfellow2016deep,
  title={Deep learning},
  author={Goodfellow, Ian and Bengio, Yoshua and Courville, Aaron and Bengio, Yoshua},
  year={2016},
  publisher={MIT press Cambridge}
}

@misc{gnettrackApp,
  author       = {{Gyokov Solutions}},
  title        = {{G-NetTrack Pro}},
  howpublished = {\url{https://www.gyokovsolutions.com/G-NetTrack%20Android.html}},
  year         = {2025}
}

@TechReport{huffordDEPARTMENTCOMMERCE,
  author      = {A. G. Longley and P. L. Rice},
  institution = {U.S. Department of Commerce, Environmental Science Services Administration, Institute for Telecommunication Sciences},
  title       = {Prediction of tropospheric radio transmission loss over irregular terrain: A computer method-1968},
  year        = {1968},
  note        = {doi: 10.70220/d3k87mv1 },
  number      = {ERL 79-ITS-67},
}

@TechReport{3gpp.38.901,
  author      = {{3GPP}},
  institution = {3rd Generation Partnership Project (3GPP)},
  title       = {Study on channel model for frequencies from 0.5 to 100 {GHz}},
  year        = {2020},
  month       = {01},
  note        = {Version 16.1.0},
  number      = {38.901},
  type        = {Technical report (TR)},
}

@TechReport{3gpp36873,
  author      = {{3GPP}},
  title       = {Study on {3D} channel model for {LTE}},
  year        = {2014},
  month       = {June},
  note        = {Release 12},
  number      = {TR 36.873},
  type        = {Technical Report},
}

@TechReport{3gpp36814,
  author      = {{3GPP}},
  title       = {{E-UTRA; Further advancements for {E-UTRA} physical layer aspects (Release 9)}},
  year        = {2010},
  month       = {March},
  number      = {TR 36.814},
  type        = {Technical Report},
  note        = {v9.2.0},

}

@inproceedings{zhang2015multi,
  title={Multi-source domain adaptation: A causal view},
  author={Zhang, Kun and Gong, Mingming and Sch{\"o}lkopf, Bernhard},
  booktitle={Proceedings of the AAAI Conference on Artificial Intelligence},
  volume={29},
  number={1},
  year={2015}
}

@techreport{itu2017sm2028,
  author       = {{International Telecommunication Union}},
  title        = {{Report ITU-R SM.2028-2: Monte Carlo simulation methodology for the use in sharing and compatibility studies between different radio services or systems}},
  institution  = {International Telecommunication Union},
  address      = {Geneva, Switzerland},
  year         = {2017},
  month        = {June},
  type         = {Technical Report},
  number       = {SM.2028-2}
}

@inproceedings{vanleer2021improving,
  title={Improving Propagation Model Predictions via Machine Learning with Engineered Features},
  author={Vanleer, Ann and Anderson, Christopher R},
  booktitle={MILCOM 2021-2021 IEEE Military Communications Conference (MILCOM)},
  pages={420--425},
  year={2021},
  organization={IEEE}
}

@inproceedings{kortylewski2019analyzing,
  title={{Analyzing and reducing the damage of dataset bias to face recognition with synthetic data}},
  author={Kortylewski, Adam and Egger, Bernhard and Schneider, Andreas and Gerig, Thomas and Morel-Forster, Andreas and Vetter, Thomas},
  booktitle={Proceedings of the IEEE/CVF Conference on Computer Vision and Pattern Recognition Workshops},
  year={2019}
}

@inproceedings{kozma2023proposed,
  title={A Proposed Mid-band Statistical Clutter Propagation Model Utilizing Lidar Data},
  author={Kozma, William and Cotton, Michael},
  booktitle={2023 17th European Conference on Antennas and Propagation (EuCAP)},
  pages={1--5},
  year={2023},
  organization={IEEE}
}

@article{dempsey2025reciprocity,
  title={Reciprocity-Aware Convolutional Neural Networks for Map-Based Path Loss Prediction},
  author={Dempsey, Ryan G and Ethier, Jonathan and Yanikomeroglu, Halim},
  journal={arXiv preprint arXiv:2504.03625},
  year={2025}
}

@article{akrout2023domain,
  title={Domain generalization in machine learning models for wireless communications: Concepts, state-of-the-art, and open issues},
  author={Akrout, Mohamed and Feriani, Amal and Bellili, Faouzi and Mezghani, Amine and Hossain, Ekram},
  journal={IEEE Communications Surveys \& Tutorials},
  volume={25},
  number={4},
  pages={3014--3037},
  year={2023},
  publisher={IEEE}
}

@article{douzas2018improving,
  title={{Improving imbalanced learning through a heuristic oversampling method based on K-means and SMOTE}},
  author={Douzas, Georgios and Bacao, Fernando and Last, Felix},
  journal={Information sciences},
  volume={465},
  pages={1--20},
  year={2018},
  publisher={Elsevier}
}

@article{zhang2019propagation,
  title={Propagation modeling through foliage in a coniferous forest at {28 GHz}},
  author={Zhang, Yaguang and Anderson, Christopher R and Michelusi, Nicolo and Love, David J and Baker, Kenneth R and Krogmeier, James V},
  journal={IEEE Wireless Communications Letters},
  volume={8},
  number={3},
  pages={901--904},
  year={2019},
  publisher={IEEE}
}

@article{ayadi2015two,
  title={Two-dimensional deterministic propagation models approach and comparison with calibrated empirical models},
  author={Ayadi, Mohamed and Torjemen, Nabil and Tabbane, Sami},
  journal={IEEE Transactions on Wireless Communications},
  volume={14},
  number={10},
  pages={5714--5722},
  year={2015},
  publisher={IEEE}
}

@article{seib2020mixing,
  title={Mixing Real and Synthetic Data to Enhance Neural Network Training--A Review of Current Approaches},
  author={Seib, Viktor and Lange, Benjamin and Wirtz, Stefan},
  journal={arXiv preprint arXiv:2007.08781},
  year={2020}
}

\end{document}